\documentclass[aps,prd,superscriptaddress,nofootinbib,amsmath,amsfonts,preprintnumbers,notitlepage,10pt,english]{revtex4-1}
\usepackage{amsmath}
\usepackage{amssymb}
\usepackage{babel}

\makeatletter

\usepackage{array,multirow,graphicx}
\usepackage{dcolumn}
\usepackage{newlfont}
\usepackage{bm}
\usepackage[colorlinks,citecolor=blue,urlcolor=blue,linkcolor=blue]{hyperref}
\usepackage[figtopcap]{subfigure}
\usepackage{color}

\begin{document}

\date{\today}

\title{Uniqueness of non-trivial spherically symmetric black hole solution in  special classes of  ${\textit  F(R)}$ gravitational theory}

\author{G.G.L. Nashed}
\email{nashed@bue.edu.eg}
\affiliation {Centre for Theoretical Physics, The British University in Egypt, P.O. Box
43, El Sherouk City, Cairo 11837, Egypt}

\begin{abstract}
We show, in detail, that the only non-trivial black hole (BH) solutions for a neutral as well as a charged spherically symmetric space-times, using the class ${\textit F(R)}={\textit R}\pm{\textit F_1 (R)} $, must-have metric potentials in the form $h(r)=\frac{1}{2}-\frac{2M}{r}$ and $h(r)=\frac{1}{2}-\frac{2M}{r}+\frac{q^2}{r^2}$. These BHs have a non-trivial form of Ricci scalar, i.e., $R=\frac{1}{r^2}$ and the form of ${\textit F_1 (R)}=\mp\frac{\sqrt{\textit R}} {3M} $. We repeat the same procedure for (Anti-)de Sitter, (A)dS, space-time and got the metric potentials of neutral as well as charged in the form $h(r)=\frac{1}{2}-\frac{2M}{r}-\frac{2\Lambda r^2} {3} $ and $h(r)=\frac{1}{2}-\frac{2M}{r}+\frac{q^2}{r^2}-\frac{2\Lambda r^2} {3} $, respectively. The Ricci scalar of the (A)dS space-times has the form ${\textit R}=\frac{1+8r^2\Lambda}{r^2}$ and the form of ${\textit F_1(R)}=\mp\frac{\textit 2\sqrt{R-8\Lambda}}{3M}$. We calculate the thermodynamical quantities, Hawking temperature, entropy, quasi-local energy, and Gibbs-free energy for all the derived BHs, that behaves asymptotically as flat and (A)dS, and show that they give acceptable physical thermodynamical quantities consistent with the literature. Finally, we prove the validity of the first law of thermodynamics  for those BHs.
\end{abstract}


\maketitle
\section{\bf Introduction}
To describe the early and late cosmic epoch of our universe consistently, scientists invented the modified gravitational theories that contain higher-order curvature expressions. Such expressions are responsible to make the theories renormalizable  which means that they are quantizable theories of gravitation \cite{PhysRevD.16.953}. Moreover, such theories are interesting to teach us how to understand the presence of dark matter and confront such theories with observation \cite{Nojiri:2006ri,Copeland:2006wr,Nojiri:2010wj,Awad:2017ign,Clifton:2011jh,Tang:2019qiy,Awad:2017yod}. At the beginning of the formulation of modified gravitational theories, different constructions have been done that involve all the second-order curvature scalar \cite{Podolsky:2018pfe,Lu:2015psa,Bueno:2017sui,Bueno:2016lrh}. Additionally, there was a particular class that contains the higher-order curvature invariants that constructed from the Ricci scalar \cite{DeFelice:2010aj,Cognola:2007zu,Cognola:2007zu,PhysRevD.77.023503,PhysRevD.81.049901,Zhang:2005vt,Li:2007xn,Song:2007da,Nojiri:2007cq,
Nojiri:2007as,Capozziello:2018ddp,Vainio:2016qas,Ostrogradsky:1850fid}. In spite that ${\textit F (R)} $ theories prevent the existence of any other invariants except Ricci scalar they can also prevent the existence of Ostrogradskis instability \cite{Ostrogradsky:1850fid} that is a feature of higher derivative theories \cite{Woodard:2006nt}.

The simplest modification of general relativity (GR) is to include $R^d$, $d>0$, to Einstein Hilbert action and the output field equations are able to discuss the inflationary epoch \cite{Starobinsky:1980te}. Also, we can consider the term  $R^d$, $d<0$, which can  discuss the behavior of the universe at the late epoch \cite{Carroll:2003st,Carroll:2003wy,Capozziello:2002rd,Capozziello:2003gx}. In ${\textit  F(R)}$ gravitational theory, one can reproduce BHs that are compatible or different from GR \cite{Multamaki:2006zb,2018EPJP..133...18N,Multamaki:2006ym,delaCruzDombriz:2009et,2018IJMPD..2750074N,Hendi:2011eg,Nashed:2018piz,Sebastiani:2010kv}. A spherically symmetric BHs which are different from GR have been derived form a special class of ${\textit  F(R)}$, i.e., ${\textit  F(R)=R-2\alpha\sqrt{R}}$ \cite{Nashed:2019tuk,Elizalde:2020icc,Nashed:2019yto}. These BHs have  non-trivial Ricci scalar that has the form $\frac{1}{r^2}$.  In this study, we are going to prove that such BHs are  unique for the special class  ${\textit  F(R)=R\pm{\textit F_1( R)}}$.

In (2019) the event horizon telescope picked up the first image of a BH, at the center of the galaxy Messier 87 \cite{Akiyama:2019cqa}. Since then the interest in BH studies have increased rapidly. This image was a real test of  the Einstein GR theory. Moreover, the image of BH brings observations closer to the event horizon in contrast to what has been done before as a test of GR by looking at the motions of stars and gas clouds near the edge of a BH. The picked image confirmed a  dark shadow-like region, caused by gravitational bending and capture of light which was predicted by Einstein GR \cite{Falcke:1999pj}. In this study, we show that our BH solution is a unique  correction to GR modulating the shape and the size of the event horizon. By a fine measurement, improving the precision, one could discriminate between GR and $f(R)$ BH solutions. For example, if we have a corrected BH solution of GR, like what we have done in this study,  discrimination between the two models, GR and $f(R)$, could be available. We would stress that the precision of shadow measurements could select the theory.

The plan of this study is  the following. In Sec. \ref{S22} a summary of  Maxwell-${\textit  F(R)}$ gravity is given. In Sec. \ref{S33}, restricting to spherically  symmetric space-time, exactly neutral as well as charged  BH solutions of the field equations of  ${\textit  F(R)}$  are derived in detail. In Sec. \ref{S44}, the same calculations are performed for the neutral and charged cases for which the field equations contain a cosmological constant, and  BH solutions are derived. These BHs behave asymptotically as  (A)dS space-time. In Sec. \ref{S55}, we calculate the thermodynamic  quantities like entropy, Hawking temperature, and quasi-local energy to all  BHs derived in Sec. \ref{S33} and Sec. \ref{S44} and show that all the resulting thermodynamic  quantities are physically acceptable\footnote{Physical acceptable results mean that they do not contradict the previous results and at the same time have no contradiction with observation, for example, the entropy has a positive value, the temperature has a positive value, and so on.}. Finally, we show in Sec. \ref{S55} that the first law of thermodynamics is satisfied for the BHs derived in Sec. \ref{S33} and Sec. \ref{S44}. In the final section, Sec. \ref{S77}, we discuss the main results of the present study,  figured out some convincing conclusions, and investigate some ideas for future study.

\section{A brief summary of the Maxwell--${\textit  F(R)}$ theory}\label{S22}
 ${\textit  F(R)}$ gravitational theory is a modification  of GR that is studied early by many authors (cf. \cite{Carroll:2003wy,1970MNRAS.150....1B,Nojiri:2003ft,Capozziello:2003gx,Capozziello:2011et,Nojiri:2010wj,Nojiri:2017ncd,Capozziello:2002rd}). The Lagrangian of Maxwell--${\textit F(R)}$ theory  takes the form:
\begin{eqnarray} \label{a2} {\mathop{\mathcal{ L}}}_g:=\frac{1}{2\kappa} \int d^4x \sqrt{-g} \Big[{\textit F(R)}-\Lambda-\frac{1}{2}\mathrm{f}^{2}\Big]\equiv\frac{1}{2\kappa} \int d^4x \sqrt{-g} \Big[{\textit R+F_1(R)}-\Lambda-\frac{1}{2}\mathrm{f}^{2}\Big],\end{eqnarray}
where ${\textit  R}$ is the Ricci scalar, $\kappa$ is the gravitational constant,  $\Lambda$ is the cosmological constant,  $g$ is the determinant of the metric, and ${\textit  F_1(R)}$ is an arbitrary function that can take any form. Equation (\ref{a2}) shows that when ${\textit  F_1(R)}\neq 0$ we have a theory  deviates from GR.  In Eq. (\ref{a2}),  $\mathrm{f}^2=\mathrm{f}_{\mu \nu}\mathrm{f}^{\mu\nu}$ and $\mathrm{f}_{\mu \nu} =\mathrm{A}_{\mu, \nu}-\mathrm{A}_{\nu, \mu}$,  where $\mathrm{A}_\mu$ is the gauge potential and the comma means ordinary differentiation.

Making  variations of the Lagrangian (\ref{a2}) w.r.t. the metric tensor $g_{\mu \nu}$ and the strength tensor $\mathrm{f}$, respectively, we can write the  field equations of the Maxwell--${\textit F(R)}$ gravitational  theory as  \cite{2005JCAP...02..010C}
\begin{eqnarray} \label{f1}
{\textit E}_{\mu \nu}={\textit  R}_{\mu \nu} \Big[{\textit 1+F_1(R)}\Big]-\frac{1}{2}g_{\mu \nu}\Big[{\textit R+F_1( R)}\Big]-2g_{\mu \nu}\Lambda +g_{\mu \nu} \Box {\textit F}_{_{\textit 1R}}-\nabla_\mu \nabla_\nu {\textit F}_{_{\textit 1R}}-8\pi {\textit T_{\mu \nu}}\equiv0,\end{eqnarray}
\begin{equation}\label{fe2}
{\partial}_{\nu} \left(\sqrt{-{\textit g}}\,{ \mathrm{f}}^{\mu \nu} \right)=0, \end{equation}
 where $\Box$ is the d'Alembertian operator, $\displaystyle {\textit F}_{_{\textit 1R}}=\frac{dF_1({\textit R})}{d{\textit R}}$  and the energy-momentum tensor, ${\textit T_{\mu \nu}}$, is defined as
 \begin{eqnarray} {\textit T_{\mu \nu}}:=\frac{1}{4\pi}\left({ \textit g}_{\rho \sigma}{{  \mathrm{f}}_\nu{}^\rho}{{{\mathrm{f}}}_\mu}^{\sigma}-\displaystyle{1 \over 4}  {\textit g}_{\mu \nu} \mathrm{f}^{2}\right).\end{eqnarray}
The trace of  Eq.~(\ref{f1}), takes the form
\begin{eqnarray} \label{f3}
{\textit E}={\textit R}{\textit F}_{\textit 1R}-2{\textit F(R)}-8\Lambda+3\Box {\textit F}_{\textit 1R}=0 \,.\end{eqnarray}
Equation (\ref{f3}) has no contribution from the electromagnetic part due to the skewness of this part.  In the following section,  we are going to apply the field equations (\ref{f1}),  (\ref{fe2}) and (\ref{f3}), with/without charged and with/without  cosmological constant for a particular form of spherically symmetric space-time.
\section{Black holes with asymptotic flatness}\label{S33}
In this section, we are going to apply the neutral and charged  field equations (\ref{f1}),  (\ref{fe2}) and (\ref{f3}) without the cosmological constant to a spherically symmetric space-time and try to derive a general form of the arbitrary function ${\textit F_1(R)}$.
\subsection{A neutral  BH with asymptote flat space-time}
It is well known that in the frame of GR any neutral spherically symmetric solution of the metric
\begin{eqnarray} \label{met12}
& &  ds^2=-h(r)dt^2+\frac{dr^2}{h(r)}+r^2(d\theta^2+\sin^2d\phi^2),  \end{eqnarray}
has  the following form \begin{eqnarray}\label{met122} h(r)=1-\frac{2M}{r}\, , \end{eqnarray} where $M$ is the gravitational mass of the BH. Equation  (\ref{met12}) with Eq. (\ref{met122})  has a vanishing Ricci scalar, i.e., ${\textit R}=0$, and therefore can be a solution to ${\textit F(R)}$ gravitational theory\footnote{This solution is known as the Schwarzschild solution \cite{1916AbhKP1916..189S}.}. However, when  \begin{eqnarray} \label{met2} h(r)=\frac{1}{2}-\frac{2M}{r}\,, \end{eqnarray}  the Ricci scalar becomes not trivial, i.e., ${\textit R}=\frac{1}{r^2}$ and the metric (\ref{met12}) will be a solution to the class ${\textit F(R)}={\textit R}-2\alpha \sqrt{{\textit R}}$ where $M=\frac{1}{6\alpha}$ \cite{Nashed:2019tuk}. In this section, we are going to search if there is any other non-trivial solution  in the frame of ${\textit F(R)}$ gravitational theory. For this purpose we assume that  the metric potential to have the form  \begin{eqnarray} \label{met33} h(r)=\xi-\frac{2M}{r}\,, \end{eqnarray} where $\xi$ is an arbitrary real parameter. Equation (\ref{met33}) coincides with Eqs. (\ref{met12}) and (\ref{met2}) when $\xi=1$ and $\xi=1/2$ respectively.
  The Ricci scalar of the metric (\ref{met12}) using Eq. (\ref{met33}) has the form
  \begin{eqnarray} \label{Ricci}
  {\textit R(r)}=\frac{2(1-\xi)}{r^2}\Rightarrow{\textit r(R)}=\pm\frac{2(1-\xi)}{\sqrt{\textit R}}.
  \end{eqnarray}
  Applying  the field equations
 (\ref{f1}) and (\ref{f3}) to  Eq. (\ref{met12}) using the metric potential (\ref{met33}) we get the following system of non-linear differential equations
 \begin{eqnarray}
&& {\textit E}_t{}^t=\frac {1}{2r^2}\left( 2\,\xi-2-\mathbb{F}{r}^{2}
 \right) +\frac{{r}^{2}\left( r\xi-2\,M \right) \mathbb{F}'''-{r} \left( 15\,M-8\,{r}\xi
 \right)\mathbb{F}'' - 3\left(7\,M-4\,{r}\xi \right)\mathbb{F}'}{ 4\left( \xi-1 \right)}=0\,,\nonumber\\
&&{\textit E}_r{}^r=\frac {1}{2r^2} \left( 2\,\xi-2-\mathbb{F}{r}^{2} \right) -\frac {{r}\left( 3\,M-2\,{r}\xi \right) \mathbb{F}''+3 \left( 3\,M-2\,{r}\xi
 \right)\mathbb{F}'  }{4\left( \xi-1
 \right) }=0\,,\nonumber\\
&&{\textit E}_\theta{}^\theta=E_\phi{}^\phi=-\frac {\mathbb{F}}{2}+\frac { {r}^{2}\left( {r}\xi-2\,M \right) \mathbb{F}'''-r\left( 12\,M-7\,\xi\,{r}\right)\mathbb{F}'' + \left[  \left( 8\,\xi+1 \right)
r-12\,M \right]\mathbb{F}'}{4(\xi-1)}=0\,,\nonumber\\
&&{\textit E}=\frac {2}{r^2} \left(\xi-1-\mathbb{F}  {r}^{2} \right) + \frac {3{r}^{2}\left( \,{r}\xi-2\,M \right)\mathbb{F}'''-6{r}\left( 7\,M-4\,{r}\xi
 \right)\mathbb{F}''+ 2\left[ \left(
17\,\xi+1\right) {r}-27\,M \right]\mathbb{F}'
 }{4\left( \xi-1 \right)}=0\,,
\label{feq}
\end{eqnarray}
where $\mathbb{F}'=\frac{d\mathbb{F}(r)}{dr}$, $\mathbb{F}''=\frac{d^2\mathbb{F}(r)}{dr^2}$, $\mathbb{F}'''=\frac{d^3\mathbb{F}(r)}{dr^3}$, $ {\textit F_1( R(r))}=\mathbb{F}\equiv \mathbb{F}(r)$ and  $\frac{\textit dF_1( R(r))}{d \textit ( R(r))}=\frac{d\mathbb{F}(r)}{dr}\frac{dr}{d \textit ( R(r))}$. Equation (\ref{feq}) informs us that  when $\mathbb{F}=0$ then  $\xi=1$.  In that case, $\mathbb{F}=0$,  there will be no new solution different from the  well-known solution of GR as it should be  because ${\textit R}=0$ as is clear from Eq. (\ref{Ricci}). Assuming that $\mathbb{F}\neq 0$ and  $\xi\neq 1$ we get the  following two solutions of the system (\ref{feq}):
  \begin{eqnarray} \label{soll1}
 && \xi=1+\frac{c_1}{2}, \qquad \qquad {\textrm and} \qquad \qquad \mathbb{F}(r)=\frac{c_1}{r^2}\,,\nonumber\\
&&\xi=\frac{1}{2}, \qquad \qquad \qquad {\textrm and} \qquad \qquad \mathbb{F}(r)=\frac{1}{3Mr}\,.
   \end{eqnarray}
Using Eq. (\ref{Ricci}) in the first set of Eq. (\ref{soll1}) we get
\begin{eqnarray} \label{sol1}
{\textit F(R)} \propto {\textit R} \qquad \qquad {\textrm which\, \, is\, \, the\, \, case\, \, of\, \, GR}
\end{eqnarray}
while if we use Eq. (\ref{Ricci}) in the second set of Eq. (\ref{soll1}) we get
\begin{eqnarray} \label{sol11f}
{\textit F(R)}={\textit R}\mp\frac{\sqrt{\textit R}}{3M}.
\end{eqnarray}
Equation (\ref{sol11f}) shows that for the metric potential (\ref{met33}) {\textrm the only non-charged BH solution that can  deviates from GR comes from the contribution of ${\textit F_1(R)}$ that has the only following form}
\begin{eqnarray} \label{sol111}
{\textit F_1(R)}=\mp\frac{\sqrt{\textit R}}{3M},
\end{eqnarray}
where $\xi=1/2$. Equation (\ref{sol111}) coincides  with that derived in \cite{Nashed:2019tuk,Nashed:2019yto} when $M=\frac{1}{6\alpha}$.  Now we are going to study the charged form of ${\textit F(R)}$ and see what is the form of ${\textit F_1(R)}$ in that case.
\subsection{ Charged BH with asymptotic flatness}
It is well-know that the charged BH solution in the frame of GR has a metric potential in the form
 \begin{eqnarray}\label{met11} h(r)=1-\frac{2M}{r}+\frac{q^2}{r^2}\, , \end{eqnarray} where $q$ is the charge parameter which comes from the electromagnetic potential $\mathrm{A}$ that comes from the definition $\mathrm{f}_{\mu \nu}=\mathrm{A}_{\mu,\nu}-\mathrm{A}_{\nu,\mu}$  where the vector $\mathrm{A}_\mu$  has the form $\mathrm{A}=(q/r,0,0,0)$ \cite{1916AnP...355..106R}.  Equation (\ref{met11}) has a vanishing Ricci scalar. However, when  \begin{eqnarray} \label{met22} h(r)=\frac{1}{2}-\frac{2M}{r}+\frac{q^2}{r^2}\,, \end{eqnarray}  the Ricci scalar is not trivial and takes the form $R=\frac{1}{r^2}$ and the metric (\ref{met11}) will be a solution to the class ${\textit F(R)}={\textit R}-2\alpha \sqrt{{\textit R}}$ \cite{Nashed:2019tuk}. Now, assuming  the metric potential to have the form  \begin{eqnarray} \label{met333f} h(r)=\xi_1-\frac{2M}{r}+\frac{q^2}{r^2}\,, \end{eqnarray} where $\xi_1$ is an arbitrary  real parameter.   Applying  the field equations
 (\ref{f1}), (\ref{fe2}) and (\ref{f3}) to  Eq. (\ref{met12}) using the metric potential (\ref{met333f})  we get the following system of non-linear differential equations
 \begin{eqnarray}
&& {\textit  E}_t{}^t=\frac {1}{2r^2}\left( 2\,\xi-2-\mathbb{F}{r}^{2}
 \right)+\frac{ \left(\xi_1 -\frac{2\,M}{r}+\frac{{q}^{2}}{r^2}\right){r}^{3} \mathbb{F}''' + \left(8\,\xi_1-\frac{15\,M}{r}+7\,\frac{{q}^{2}}{r^2} \right){r}^{2}\mathbb{F}'' + \left( 8\,\frac{{q}^{2}}{r^2}-\frac{21\,M}{r}+12\,\xi_1\,
 \right){r}\mathbb{F}' }{ 4\left( \xi_1-1 \right)}
=0\,,\nonumber\\
&&{\textit  E}_r{}^r=\frac {1}{2r^2}\left( 2\,\xi-2-\mathbb{F}{r}^{2}
 \right)+\frac { \left(2\,\xi_1 -\frac{3\,M}{r}+\frac{{q}^{2}}{r^2}
 \right){r}^{2}\mathbb{F}'' + \left( 6\,\xi_1-\frac{9\,M}{r}+2\,\frac{{q}^{2}}{r^2} \right){r}\mathbb{F}'}{4
 \left( \xi_1-1 \right)}
=0\,,\nonumber\\
&&{\textit  E}_\theta{}^\theta=E_\phi{}^\phi=\frac{-\mathbb{F}}{2}+\frac { \left(\xi_1 -\frac{2\,M}{r}+\frac{{q}^{2}}{r^2}\right) {r}^{3}
\mathbb{F}'''+ \left(7\,\xi_1-\frac{12\,M}{r}+5\,\frac{{q}^{2}}{r^2} \right){r}^{2}\mathbb{F}''+ \left(  \left( 8\,\xi_1+1 \right) -\frac{12\,M}{r}+4\,\frac{{q}^{2}}{r^2} \right){r
}\mathbb{F}' }{
 4\left( \xi_1-1 \right)}
=0\,,\nonumber\\
&&{\textit  E}=\frac {2}{r^2} \left(\xi-1-\mathbb{F}  {r}^{2} \right) +\frac { 3\left(\xi_1 -\frac{2\,M}{r}+\,\frac{{q}^{2}}{r^2}
 \right){r}^{3}\mathbb{F}''' +6\left(4\,\xi_1-\frac{7\,\,M}{r}+3\,\frac{{q}^{2}}{r^2}\right){r
}^{2}\mathbb{F}'' +2 \left(  \left(1+17\,\xi_1 \right)-\frac{27\,M}
{r}+9\,\frac{{q}^{2}}{r^2} \right)r \mathbb{F}'}{4\left( \xi_1-1 \right) }
=0.\nonumber\\\label{feq1}
&&
\end{eqnarray}
Equations (\ref{feq1})  reduce to Eqs.  (\ref{feq})  when the charge parameter $q=0$.  Also Eqs. (\ref{feq1})  indicate that  when $\mathbb{F}=0$ we get  $\xi_1=1$ and similar  discussion carried out for the neutral BH can  also apply to the charged case.  Solving system (\ref{feq1}) assuming that $\mathbb{F}\neq 0$ and $\xi_1\neq 1$  we get the following solution
  \begin{eqnarray} \label{soll11f}
&&\xi_1=\frac{1}{2}, \qquad \qquad {\textrm and} \qquad \qquad \mathbb{F}(r)=\frac{1}{3Mr}\,.
   \end{eqnarray}
Equation (\ref{soll11f}) indicates that all the results obtained in the neutral case will be identical with  the charged case. We must stress on the fact that in the charged case the charge parameter $q$ has no effect on the form of ${\textit F_1( R)}$ because the term $q/r^2$  is a solution to the GR field equation so its effect disappears in the higher order curvature terms.

\section{ Black holes with asymptotic (ANTI-)DE-space-time}\label{S44}
 We have shown in the previous section that the only non-trivial solution, that behaves asymptotically as flat space-time, in ${\textit  F(R)}$ gravitational theory  must have a metric potential in the form $\frac{1}{2}-\frac{2M}{r}$ for the neutral case and $\frac{1}{2}-\frac{2M}{r}+\frac{q^2}{r^2}$ for the charged one. Those solutions give a non-trivial form of the Ricci scalar, ${\textit R=\frac{1}{r^2}}$, and the form of  ${\textit F_1( R)}$ is ${\textit F_1( R)}=\mp\frac{\sqrt{{\textit R}}}{3M}$.  In this section we are going to follow the same technique of the previous section and study neutral and charged BHs that behave asymptotically as (A)dS using the line-element (\ref{met12}).
\subsection{Non-charged BH with asymptotic (A)dS}
Taking  the metric potential to have the form  \begin{eqnarray} \label{met33s} h(r)=\xi_2-\frac{2M}{r}+\xi_3 r^2\,, \end{eqnarray} where $\xi_2$ and $\xi_3$ are  arbitrary real parameters.   The Ricci scalar of the metric (\ref{met12}) using Eq. (\ref{met33s}) has the form
  \begin{eqnarray} \label{Ricci1}
  {\textit R(r)}=\frac{2(1-\xi_2-6\xi_3 r^2)}{r^2}\Rightarrow{\textit r(R)}=\pm\frac{\sqrt{2(1-\xi_2)}}{\sqrt{ \textit(R+12\xi_3)}}.
  \end{eqnarray}
  Equation (\ref{Ricci1}) shows that when $\xi_2=1$ and $\xi_3=0$ we get a vanishing Ricci scalar which corresponds to a flat space-time BH solution of GR. Therefore, for non-constant Ricci scalar we must choose any value for $\xi_2$ except 1.
  Applying  the field equations
 (\ref{f1}) and (\ref{f3}) to  Eq. (\ref{met12}) using the metric potential (\ref{met33s}) we get the following system of non-linear differential equations
 \begin{eqnarray}
&& {\textit E}_t{}^t=\frac{1}{r^2}\left(\left[{ 2\Lambda} -\frac{\mathbb{F} {r}^{2}}{2}+3{\xi_3}
 \right] {r}^{2}+\xi_2-1 \right)  +\frac { 1}{
 4\left(\xi_2-1 \right)}\Bigg\{\left(r\xi_2 -{2\,M}+{r}^{3}{ \xi_3} \right)
{r}^{2}\mathbb{F}'''+ \left(9\,
{r}^{3}{\xi_3} -{15\,M}+8r\,\xi_2 \right){r}\mathbb{F}''\nonumber\\
&&
 + 3\left(4r\xi_2 -{7\,M}+4{r}^{3}{ \xi_3} \right)\mathbb{F}'\Bigg\}
=0\,,\nonumber\\
&&{\textit E}_r{}^r=\frac{1}{r^2}\left(\left[{ 2\Lambda} -\frac{\mathbb{F} {r}^{2}}{2}+3{\xi_3}
 \right] {r}^{2}+\xi_2-1 \right)  +\frac { \left(2r\xi_2 -{3\,M}+3\,{r}^{3}{\xi_3}
 \right){r}\mathbb{F}''+ 3\left(2\,{r}^{3}{\xi_3}-{3\,M}
+2r\,\xi_2 \right){r}\mathbb{F}'
 }{ 4\left( \xi_2-1 \right)}=0\,,\nonumber\\
&&{\textit E}_\theta{}^\theta=E_\phi{}^\phi=  \left( 2\,{\Lambda}-\frac{\mathbb{F}}{2}+3{\xi_3} \right)  + \frac{1}{4\,(\xi_2-1)} \Bigg\{\left(r\xi_2 -{2\,M}+{r}^{3}{\xi_3} \right){r}^{2}\mathbb{F}''' + \left(9\,{r}^{3}
{\xi_3} -{12\,M}+7r\,\xi_2 \right) {r}\mathbb{F}''\nonumber\\
&& + \left( 12\,{\xi_3}\,{r}^{3}+ \left( 1+8\,\xi \right)r-{12M}\right)\mathbb{F}'\Bigg\}
=0\,,\nonumber\\
&&{\textit E}=\frac{2}{r^2}\left( \left[ \frac{3{\xi_3}}{2}+{\Lambda} \right]r^2-\mathbb{F}r^2  +\xi_2-1\right)+\frac {1}{4\left( \xi_2-1 \right)}  \Bigg\{3\left({r}^{3}{\xi_3} -{2\,M}+r\xi_2
 \right){r}^{2}\mathbb{F}'''+6\left({5\,{r}^{3}{\xi_3}} -{7\,M}+4r\xi_2 \right){r}\mathbb{F}''\nonumber\\
&& + 2\left( 21\,{r}^{3}{\xi_3}+ \left( 17\,\xi+
1\right)r-{27\,M} \right) \mathbb{F}'\Bigg\}
=0\,.\nonumber\\
&&
\label{feq3s}
\end{eqnarray}
  Equation (\ref{feq3s}) coincides with (\ref{feq}) when $\Lambda$ and $\xi_3=0$.  The above system shows that  when $\mathbb{F}=0$, the only solution is  $\xi_2=1$ and $\xi_3=-\frac{2\Lambda}{3}$ which is the solution  of GR theory.  Provided that $\mathbb{F}\neq 0$ and $\xi\neq 1$, the above system has the  following solution
  \begin{eqnarray} \label{soll1a}
  \xi_2=\frac{1}{2}, \qquad \qquad  \xi_3=\frac{2\Lambda}{3},\qquad \qquad {\textrm and} \qquad \qquad \mathbb{F}(r)=-\frac{1}{3rM}\,.
   \end{eqnarray}
   Before we continue to analyze Eq. (\ref{soll1a}) we must stress on the fact that any value of $\xi_2$ except  $\frac{1}{2}$ will not give any solution different from GR which due to the fact  that the form of $\mathbb{F}(r)\propto \frac{1}{r^2}$ which gives $\mathbb{F}(R) \propto R$.

   If we use Eq. (\ref{Ricci1}) in  Eq. (\ref{soll1a}) we get
\begin{eqnarray} \label{sol11}
{\textit F(R)}={\textit R}\mp\frac{\sqrt{\textit R+8\Lambda}}{3M}.
\end{eqnarray}
 The metric potential  (\ref{met33s})  with Eq. (\ref{sol11}) coincide with that derived in \cite{Nashed:2019tuk,Nashed:2019yto} when $M=\frac{1}{6\alpha}$.
\subsection{ Charged BH with (A)dS}
In this subsection we assume  the metric potential to have the form  \begin{eqnarray} \label{met333} h(r)=\xi_4-\frac{2M}{r}+\frac{q^2}{r^2}+\xi_5 r^2\,, \end{eqnarray} where $\xi_4$ and $\xi_5$  are arbitrary real parameters.
  Applying  the field equations
 (\ref{f1}), (\ref{fe2}) and (\ref{f3}) to  Eq. (\ref{met12}) using the metric potential (\ref{met333})  we get the following system of non-linear differential equations
 \begin{eqnarray}
&& {\textit E}_t{}^t=\left( 3{\xi_5}+2\,{\Lambda} \right)-\frac{\mathbb{F}}{2}+\frac{1}{r^2}[
\xi_4-1]+\frac {1 }{4\left(\xi-1 \right)}\Bigg\{\left({r^3}{\xi_5}-{2\,M}+r\xi_4+\frac{{
q}^{2}}{r} \right)r^2\mathbb{F}''' + \left(9r^3{\xi_5} -
{15\,M}+{8r\,\xi_4}+\frac{{7
q}^{2}}{r}\right)r\mathbb{F}''\nonumber\\
&& + \left(12r\xi_4 -{21\,M}+12\,{r}^{3}{\xi_5}+\frac{{8
q}^{2}}{r} \right)\mathbb{F}'\Bigg\}
=0\,,\nonumber\\
&&{\textit E}_r{}^r=\left(3 {\xi_5}+2{\Lambda} \right)-\frac{\mathbb{F}}{2} +\frac{1}{r^2}(\xi_4-1)+\frac { \left(3\,{r}^{3}{\xi_5} -{3\,M}+2r\,\xi_4+\frac{q^2}{r} \right){r}\mathbb{F}'' +
 \left( \frac{2q^2}{r} -{9\,M}+6\,{r}^{3}{\xi_5}+6r\xi_4
 \right) \mathbb{F}'}{4 \left( \xi_4-1
 \right)}
=0\,,\nonumber\\
&&{\textit E}_\theta{}^\theta=E_\phi{}^\phi=2\,{\Lambda}+{
3\xi_5}-\frac{\mathbb{F}}{2}+\frac {1}{
 4\left(\xi-1 \right) } \Bigg\{\left( {r}^{3}{\xi_5}-{2\,M}+r\xi_4+\frac{{
q}^{2}}{r} \right)r^2\mathbb{F}'''+ \left(7r\,\xi_4 -
{12\,M}+9\,{r}^{3}{\xi_5}+\frac{5\,{q}^{2}}{r}\right)r\mathbb{F}''\nonumber\\
&&+ \left(12 {\xi_5}r
^{3}+ ( 1+8\,\xi_4)r-{12M}+\frac{4{q}^{2}}{r} \right)\mathbb{F}'\Bigg\}
=0\,,\nonumber\\
&&{\textit E}=4\left(\left[3{\xi_5}+2{\Lambda}
 \right]-\frac{\mathbb{F}}{2}+\frac{\xi_4-1}{2r^2}\right)+\frac{1}{
 4\left(\xi_4-1 \right) }  \Bigg\{3\left(r^3{\xi_5} -{2\,M}+r\xi_4+\frac{{q}^{2}}{r}\right){r}^{2}\mathbb{F}'''+
 6\left(5r^3{\xi_5} -{7\,M}+{4r\xi}+\frac{3{
q}^{2}}{r} \right){r}\mathbb{F}'' \nonumber\\
&&+ 2\left(21r^3
{\xi_5}+ (1+17\xi_4)r -27\,M+\frac{9
{q}^{2}}{r} \right)\mathbb{F}'\Bigg\}
=0.\nonumber\\
&&
\label{feq3}
\end{eqnarray}
Equations (\ref{feq3})  reduce to Eqs.  (\ref{feq3s})  when the charge parameter $q=0$.  Also Eqs. (\ref{feq3})  indicate that  when $\mathbb{F}=0$ we get  $\xi=1$ as we previously discussed.  Solving the system (\ref{feq3}) assuming that $\mathbb{F} \neq 0$ and $\xi\neq 1$  we get the following solution
  \begin{eqnarray} \label{soll11}
&&\xi_4=\frac{1}{2}, \qquad \qquad \xi_5=\frac{-2\Lambda}{3},\qquad \qquad  {\textrm and} \qquad \qquad \mathbb{F}(r)=-\frac{1}{3Mr}\,.
   \end{eqnarray}
Equation (\ref{soll11}) indicates that all the results obtained in the neutral case will be identical with  the charged case.
The form of ${\textit F_1(R)}$ related to solution  (\ref{soll11}) is ${\textit F_1(R)}=-\frac{\textit 2\sqrt{R-8\Lambda}}{3M}$.

\section{Thermodynamics of the derived BHs}\label{S55}
In this section, we are going to study the thermodynamics quantities of the BHs given by Eqs. (\ref{met33}), (\ref{met333f}), (\ref{met33s}) and (\ref{met333}). For this aim we are going to write the definitions of the thermodynamic  quantities that we will use.
The  temperature of Hawking is given by \cite{PhysRevD.86.024013,Sheykhi:2010zz,Hendi:2010gq,PhysRevD.81.084040}
\begin{equation}\label{temp}
T_\pm = \frac{h'}{4\pi}\,,
\end{equation}
where $\pm$ means the inner and outer horizons.
The semi classical Bekenstein-Hawking entropy of the horizons is found to be
\begin{equation}\label{ent}
S_\pm =\Big(\frac{1}{4}A_\pm\Big){\textit F_R}\,,
\end{equation}
where $A_\pm$ is the area of the inner and outer horizons. The equation of entropy (\ref{ent}) is different from GR due to the existence of ${\textit F_R=\frac{dF(R)}{R}}$ and when ${\textit F_R}=1$, we get  ${\textit F_1(R)}=0$ as is clear from Eq. (\ref{a2}),  we return to GR. The quasi-local  energy is figured out as \cite{PhysRevD.84.023515,PhysRevD.86.024013,Sheykhi:2010zz,Hendi:2010gq,PhysRevD.81.084040,Zheng:2018fyn}
\begin{equation}\label{en}
E(r_\pm)=\frac{1}{4}\displaystyle{\int }\Bigg[2{\textit F_{R}}(r_\pm)+r_\pm{}^2\Big\{{\textit F}(R(r_\pm))-R(r_\pm){\textit F_{R}}(r_\pm)\Big\}\Bigg]dr_\pm.
\end{equation}
\subsection{Thermodynamics of the BHs (\ref{met33}) and  (\ref{met333f})}
\begin{figure}
\centering
\subfigure[~The metric potential of BH (\ref{met33})]{\label{fig:met}\includegraphics[scale=0.21]{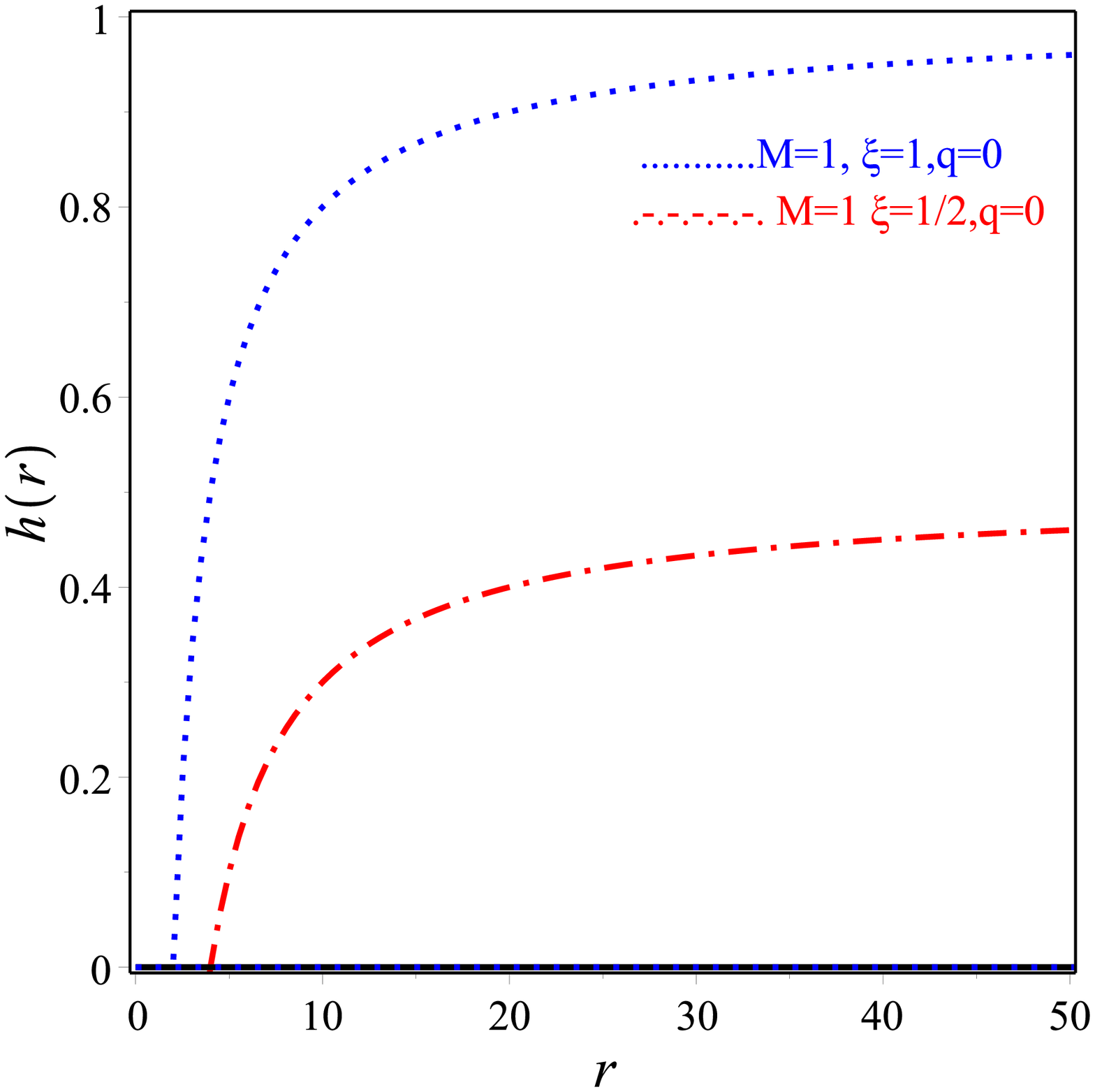}}
\subfigure[~The metric potential of BH (\ref{met333f})]{\label{fig:met1}\includegraphics[scale=0.21]{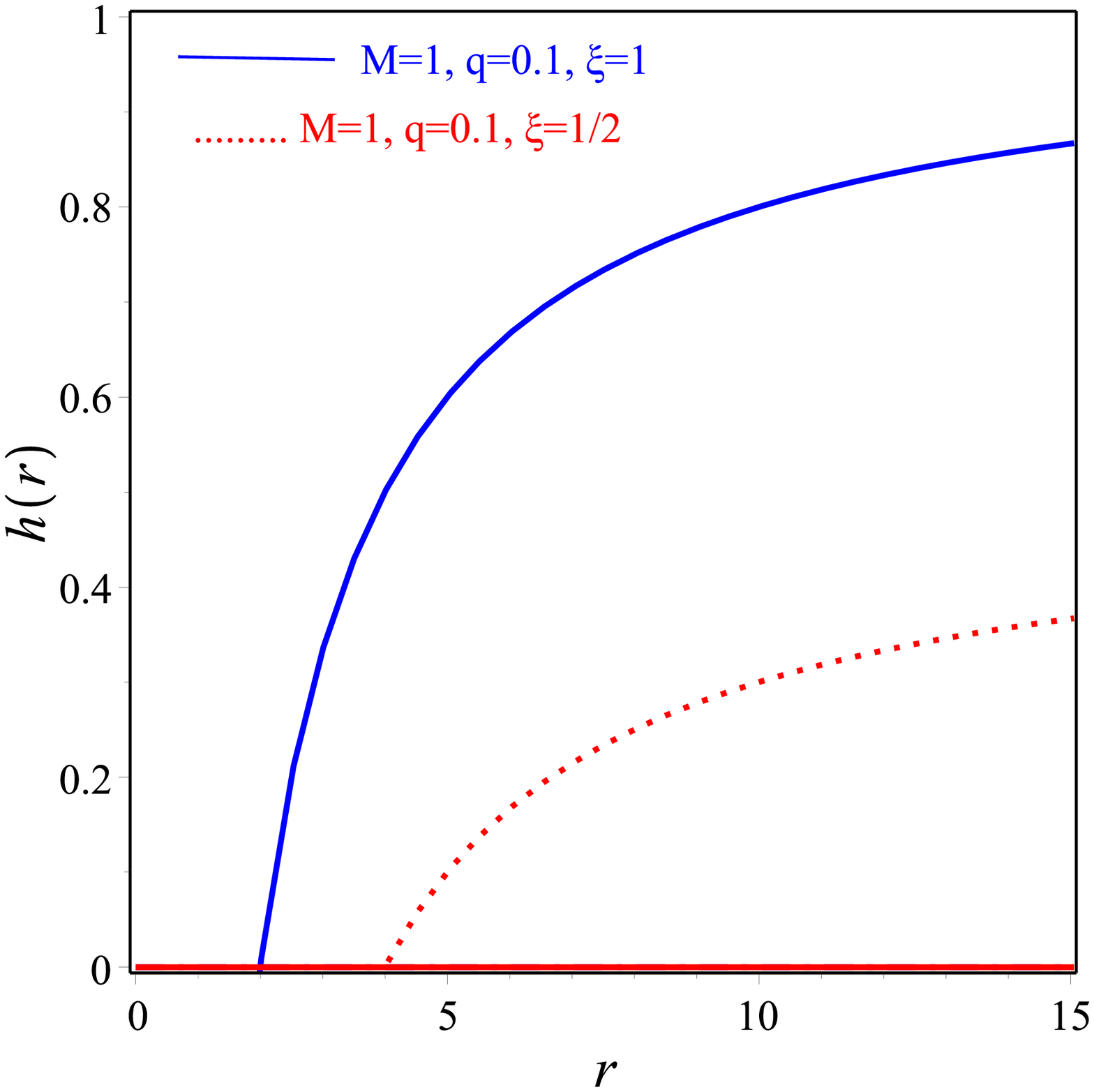}}
\subfigure[~Hawking temperature of BH (\ref{met33})]{\label{fig:Temp3}\includegraphics[scale=0.21]{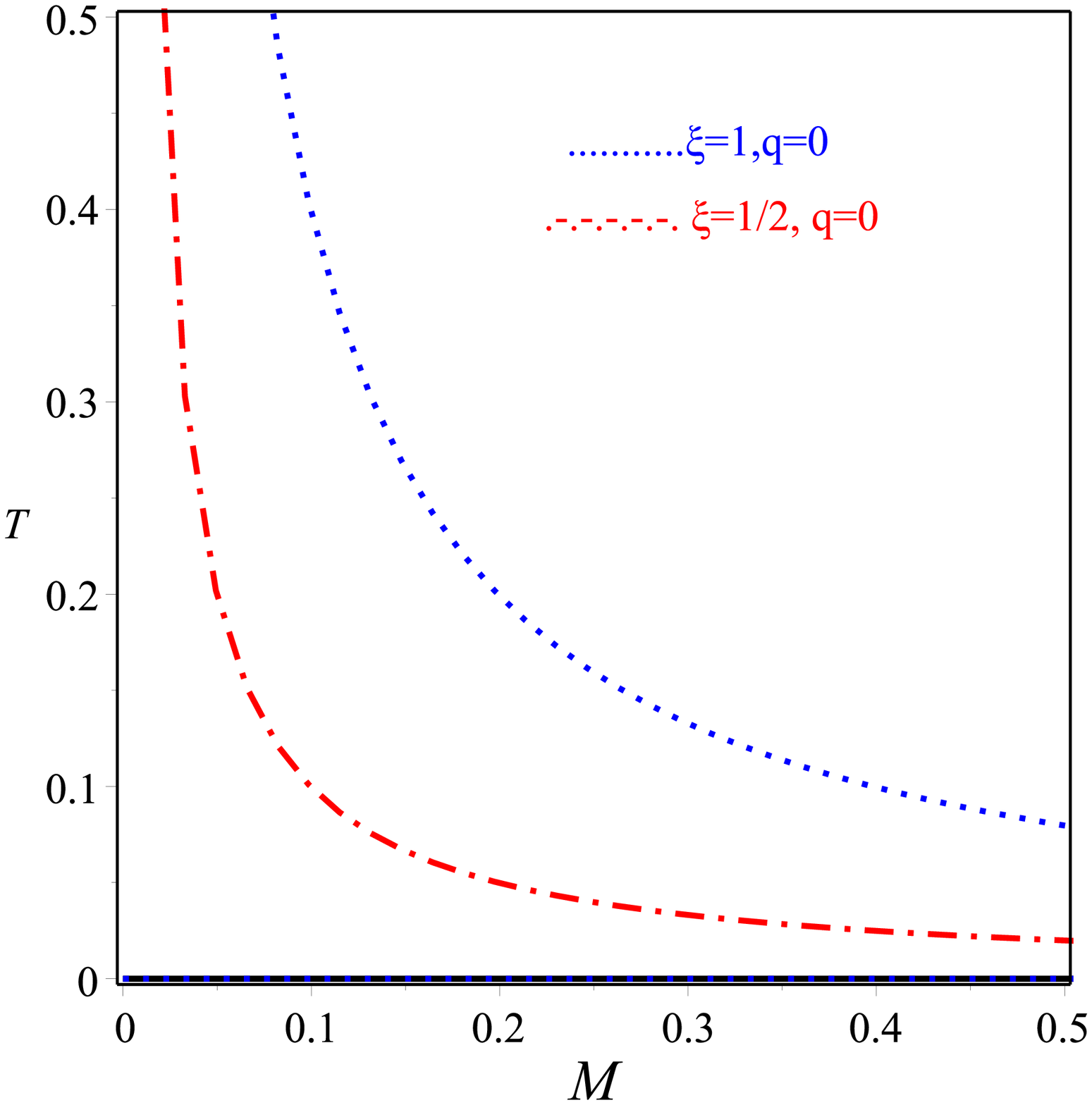}}
\subfigure[~Hawking temperature of BH (\ref{met333f})]{\label{fig:Temp1}\includegraphics[scale=0.21]{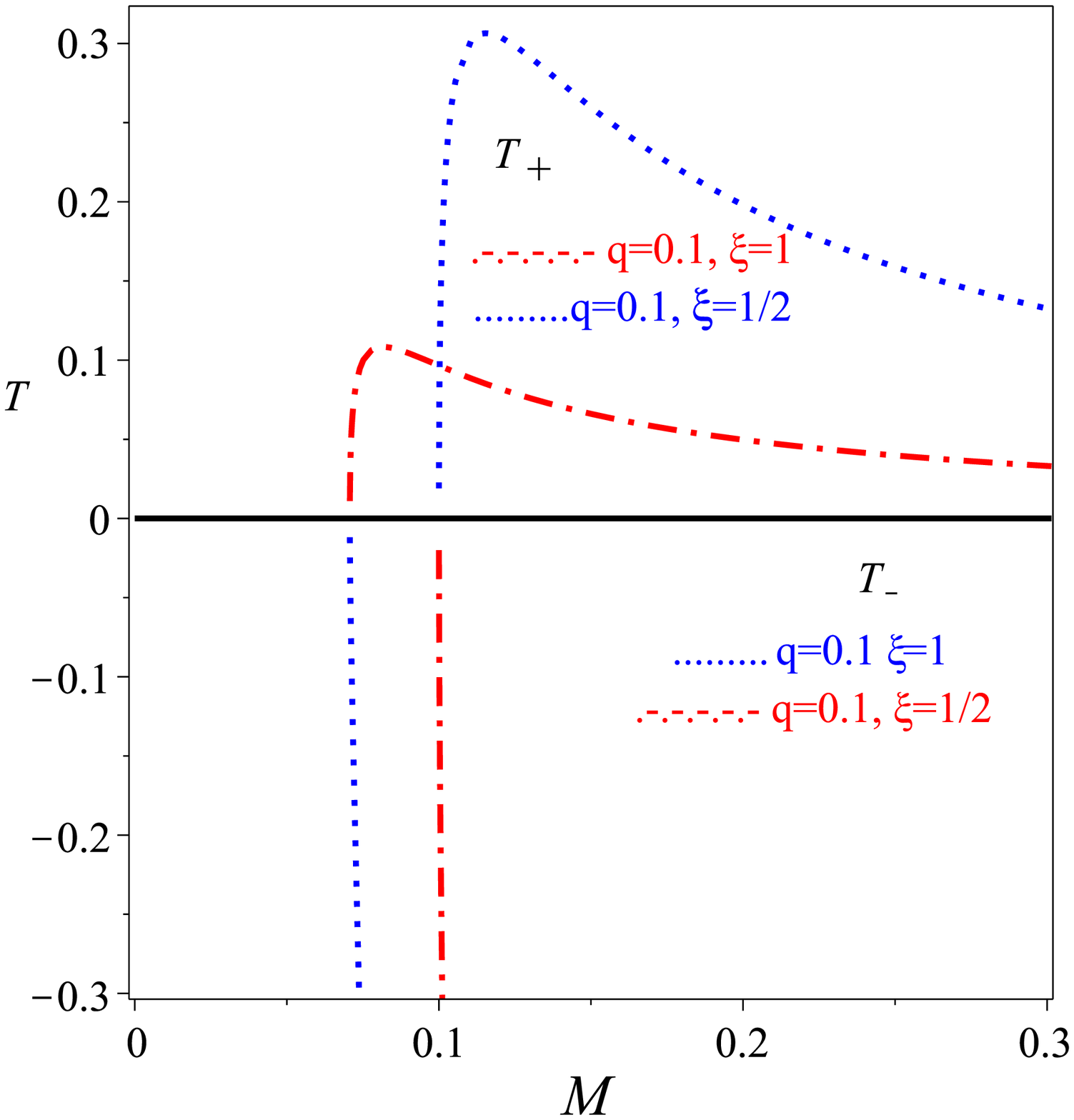}}
\subfigure[~Entropy of BH  (\ref{met33})]{\label{fig:ent3}\includegraphics[scale=0.21]{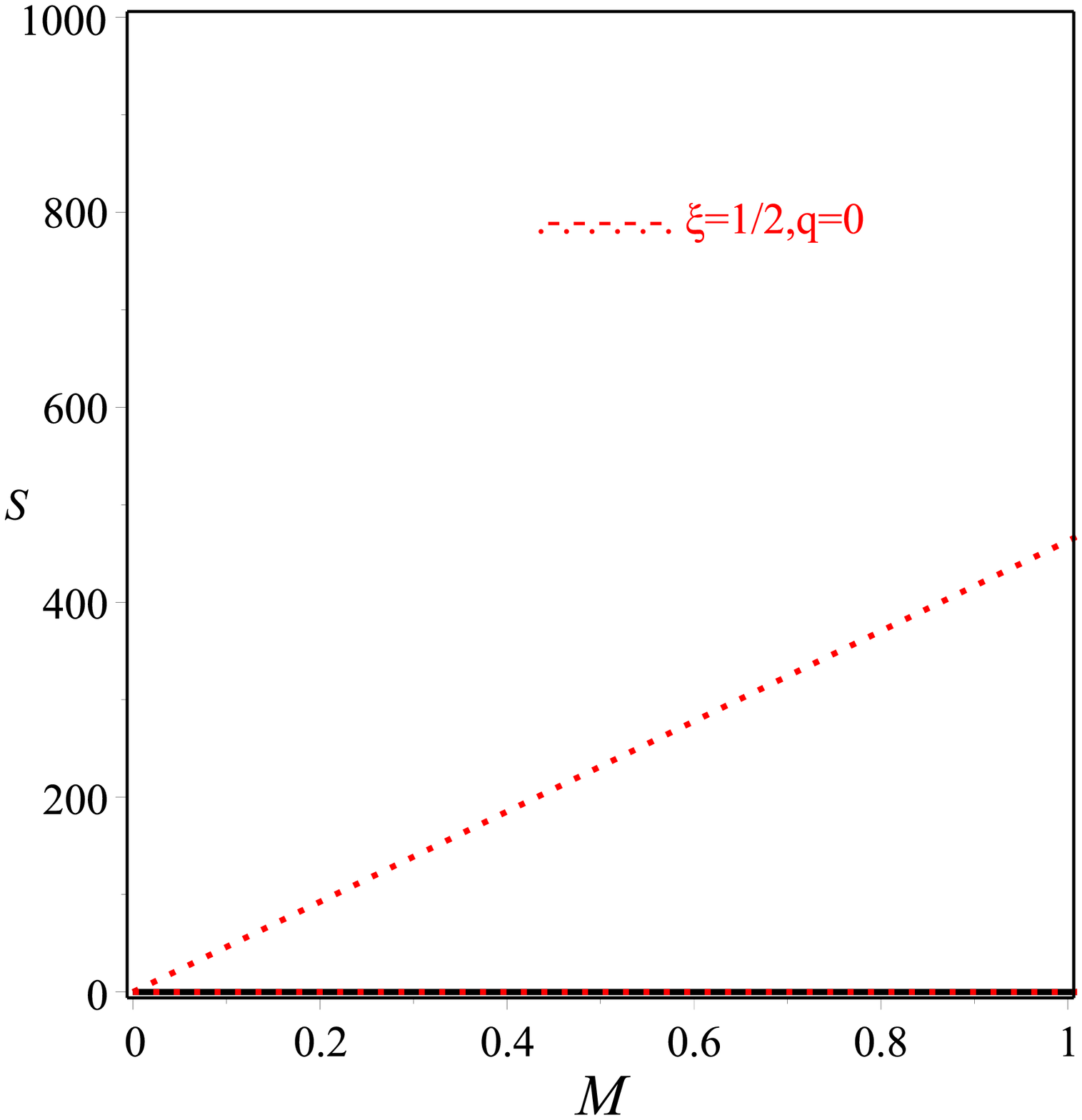}}
\subfigure[~Entropy of BH  (\ref{met333f})]{\label{fig:ent1}\includegraphics[scale=0.21]{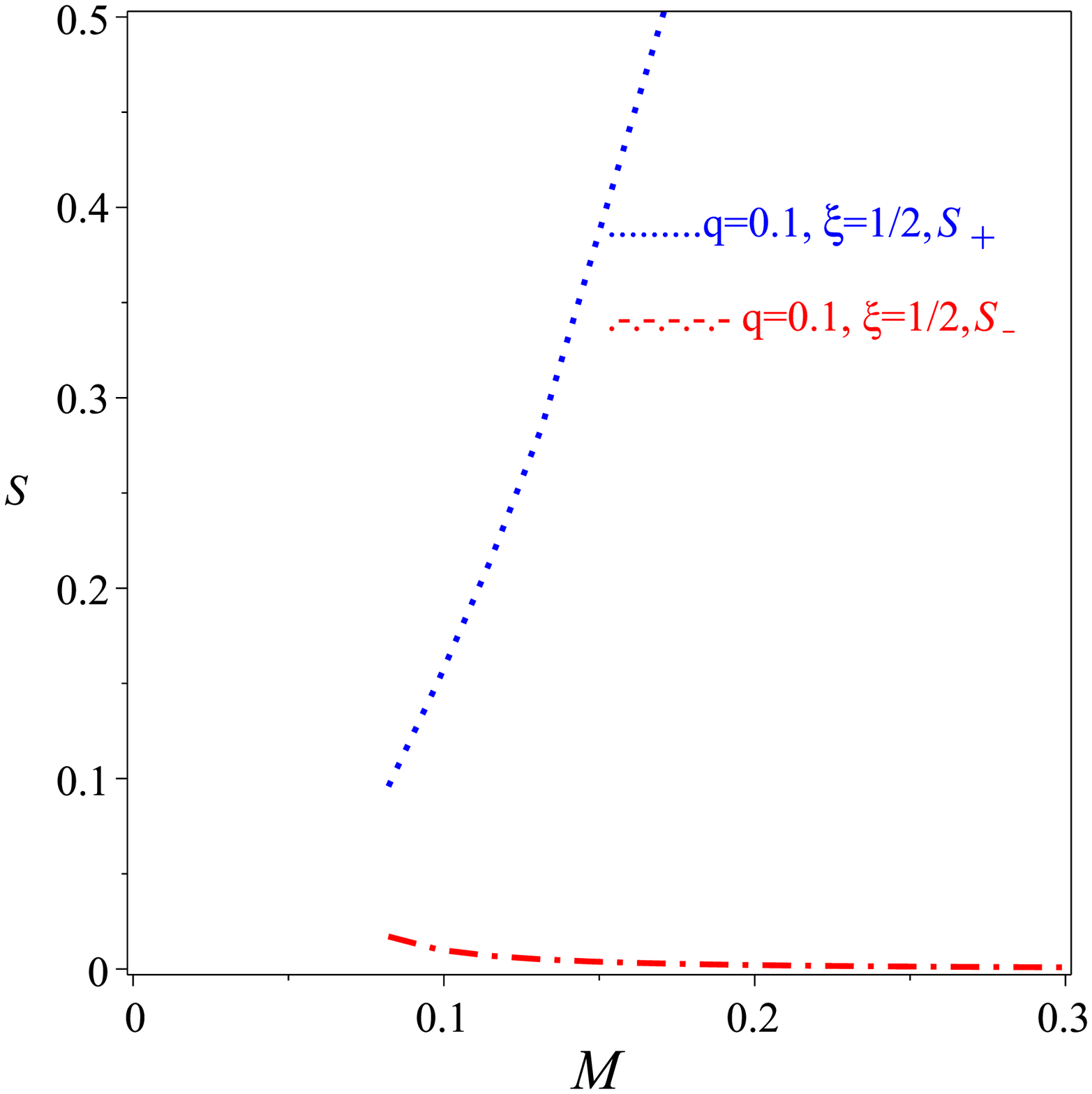}}
\subfigure[~The energy  of the BH (\ref{met33})]{\label{fig:Enr3}\includegraphics[scale=0.21]{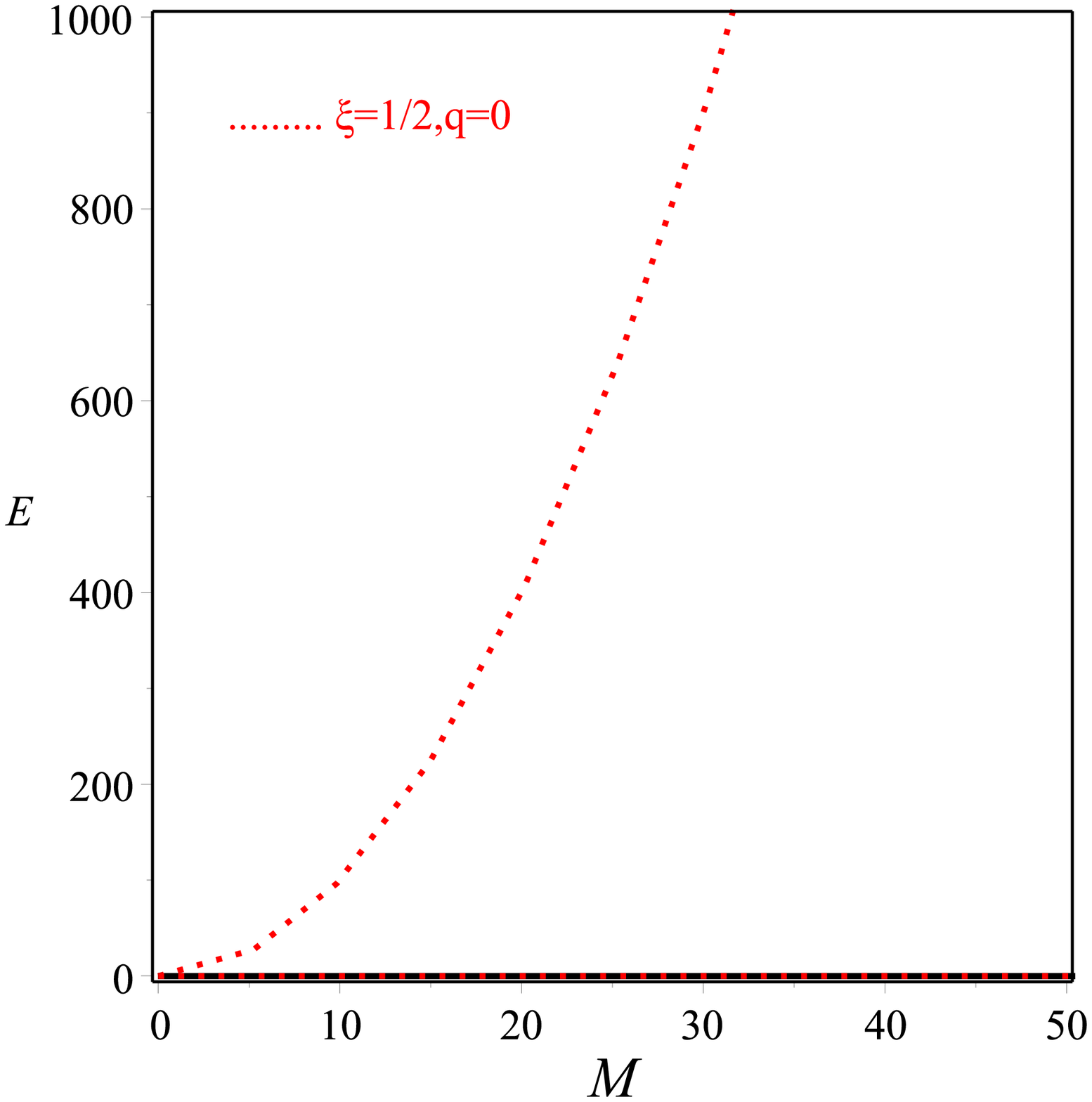}}
\subfigure[~The energy  of the BH (\ref{met333f})]{\label{fig:Enr1}\includegraphics[scale=0.21]{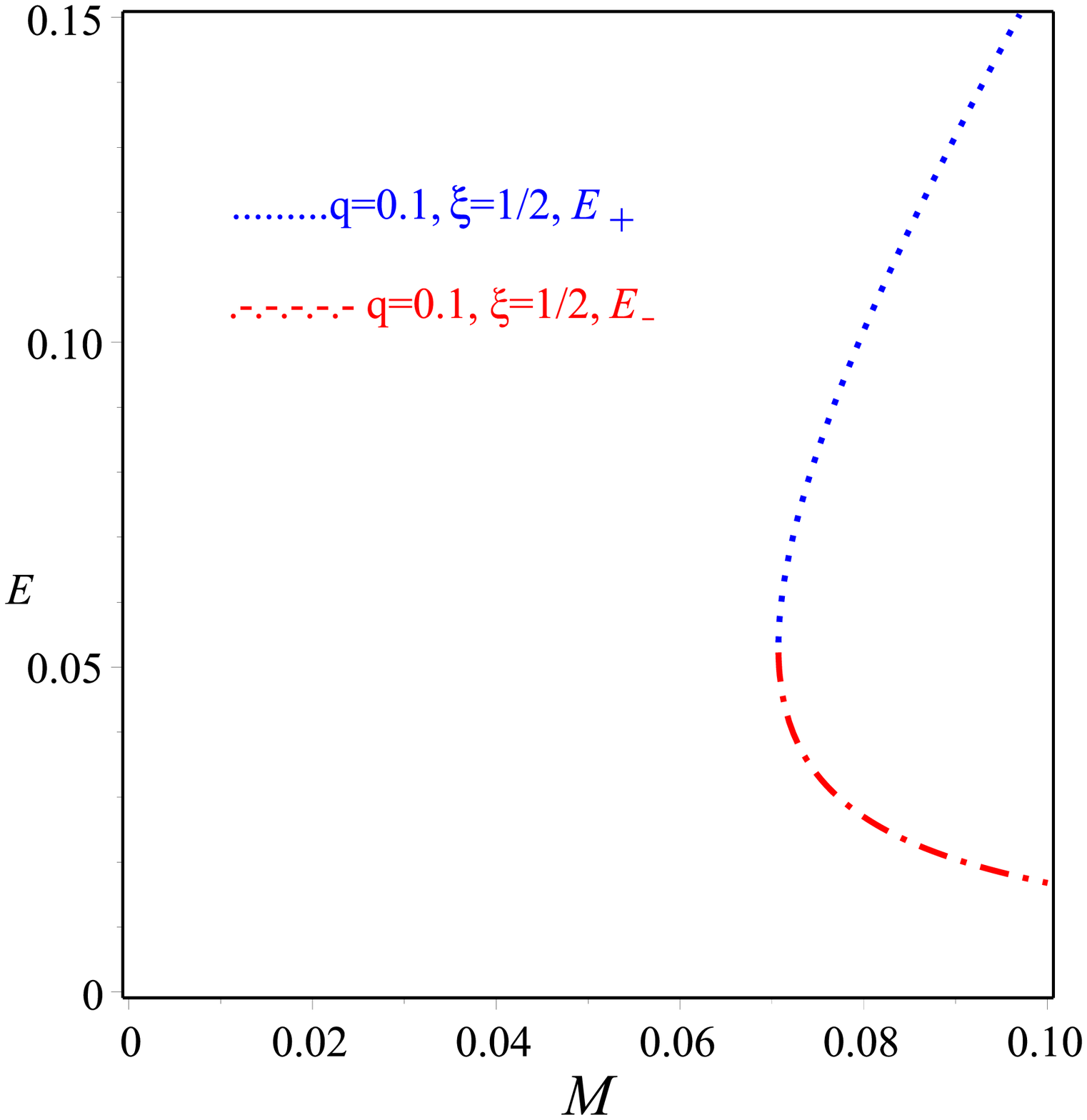}}
%
\caption[figtopcap]{\small{{Plot of the metric potential,  Hawking temperature and the entropy of  the BH  (\ref{met333f}).}}}
\label{Fig:1}
\end{figure}

The BH  (\ref{met333f})   is characterized by the  BH mass $M$, the electric charge $q$ and the parameter $\xi_1$ and when the parameter $\xi_1=1$, and $q=0$  we get the Schwarzschild space-time and when $q\neq 0$ we get Reissner-Nordstr\"om BH. The horizons of this BH have the form
\begin{eqnarray}\label{r2}
&&r_\pm=\frac{M\pm\sqrt{M^2-\xi_1 q^2}}{\xi_1}\,, \qquad \qquad M^2\geq\xi_1 q^2\,.
\end{eqnarray}
When $q=0$ we get $r=\frac{2M}{\xi_1}$ which is the horizon of the BH  (\ref{met33}). The metric potentials of the BHs  (\ref{met33}) and  (\ref{met333f}) are drawn  in Fig. \ref{Fig:1} \subref{fig:met} and \subref{fig:met1} for $\xi_1=1$ and $\xi_1=1/2$.

Using Eq. (\ref{temp}) we get the Hawking temperature of the BH (\ref{met333f}) in the form
\begin{eqnarray}\label{T2}
&&T_\pm=\frac{\xi_1{}^2(M^2\pm M\sqrt{M^2-\xi_1 q^2}-\xi_1 q^2)}{2\pi(M\pm\sqrt{M^2-\xi_1 q^2})^3}\,.
\end{eqnarray}
If we set $q=0$ in Eq. (\ref{T2}) we get $T=\frac{\xi_1{}^2}{8\pi M}$ which is the Hawking temperature of  (\ref{met33}). The behavior of the Hawking temperatures  given by Eq. (\ref{T2}) for $q=0$ and $q\neq 0$ are drawn in Fig. \ref{Fig:1} \subref{fig:Temp3} and \subref{fig:Temp1} for $\xi_1=1$ and $\xi_1=1/2$. As Fig. \ref{Fig:1} \subref{fig:Temp1}, for $q\neq 0$, shows that both values of temperatures, $\xi=1$ and $\xi=1/2$, have an increasing positive value for $T_+$ and negative values for $T_-$. Using Eq. (\ref{ent}) we get the entropy of BH (\ref{met333f})  in the form
\begin{eqnarray}\label{S2}
S_\pm=\frac{\pi(M\pm\sqrt{M^2-\xi_1 q^2})^2(12M\xi_1{}^2-12M\xi_1+M\pm \sqrt{M^2-\xi_1 q^2})}{12M\xi_1{}^3(\xi_1-1)}\,.
\end{eqnarray}
When $q=0$,  Eq. (\ref{S2}) gives $S= \frac{2\pi M^2(6\xi_1{}^2-6\xi_1-1)}{3\xi_1{}^3(\xi_1-1)}$ which is the entropy  of  the BH (\ref{met33}).
The above equation shows also that $\xi_1\neq 1$ which means that this solution cannot reduce to GR BH. The behavior of the entropy (\ref{S2}) is figured out in Fig. \ref{Fig:1} \subref{fig:ent3} and \subref{fig:ent1}, for $q=0$ and $q\neq0$.  Figure \ref{Fig:1}  \subref{fig:ent1} shows an  increasing value for $S_+$ and decreasing value for $S_-$. Using Eq. (\ref{en}) we calculate the quasi local energy and get
\begin{eqnarray}\label{E2}
E_\pm= \frac{(M\pm\sqrt{M^2-\xi_1 q^2})(24M\xi_1{}^2-M\xi_1+2M+(\pm2\mp\xi_1)\sqrt{M^2-\xi_1 q^2}-24\xi_1{}^3M)}{48\xi^2M(1-\xi_1)}\,,
\end{eqnarray}
and when $q=0$ we get $E= \frac{M(12\xi_1{}^2-\xi_1+2-12\xi_1{}^3)}{12\xi^2(1-\xi_1)}$ which is the quasi-local energy of the BH (\ref{met33}).
The behavior of the quasi local energy for $q=0$ and $q\neq 0$ are drawn  in Fig. \ref{Fig:1} \subref{fig:Enr3} and \subref{fig:Enr1} which also show  positive increasing value for $E_+$ and decreasing value for $E_-$  when $\xi_1=1/2$ and $q\neq 0$.
%
\subsection{Thermodynamics of the BHs  (\ref{met33s}) and (\ref{met333})}
\begin{figure}
\centering
\subfigure[~The metric potential of BH (\ref{met33s})]{\label{fig:met}\includegraphics[scale=0.21]{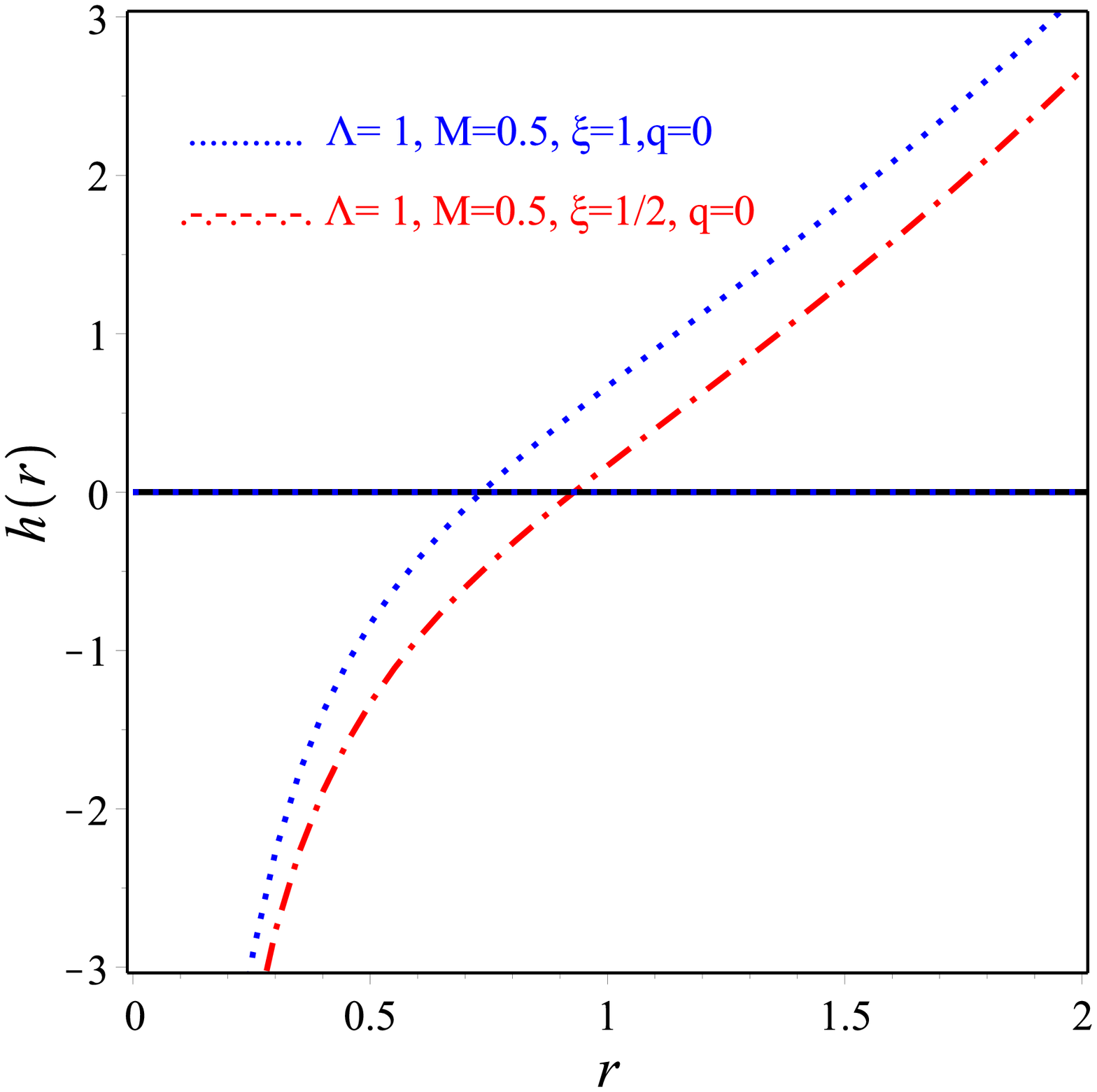}}
\subfigure[~The metric potential of BH (\ref{met333})]{\label{fig:met1}\includegraphics[scale=0.21]{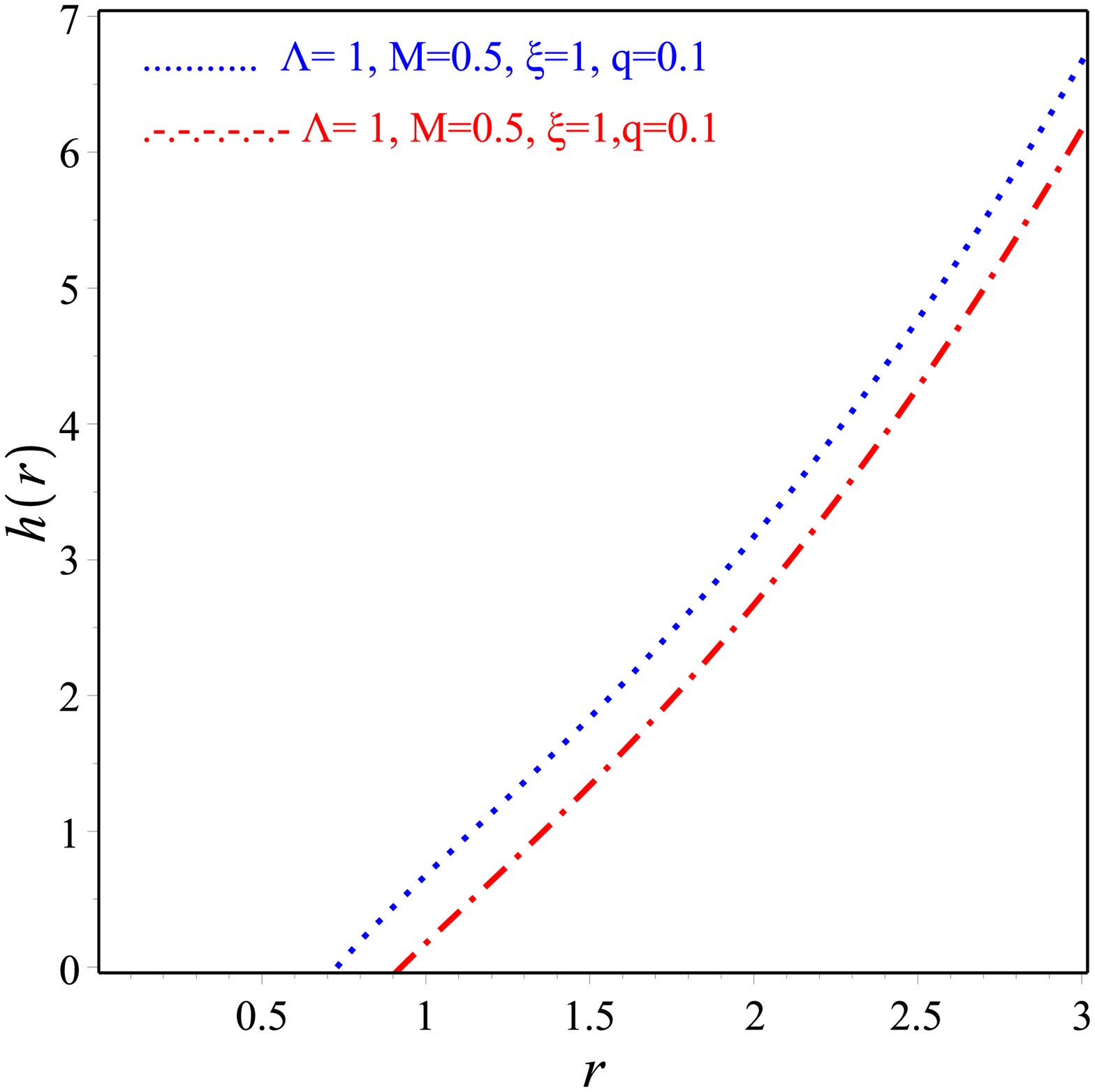}}
\subfigure[~Hawking temperature of BH (\ref{met33s})]{\label{fig:Temp3}\includegraphics[scale=0.21]{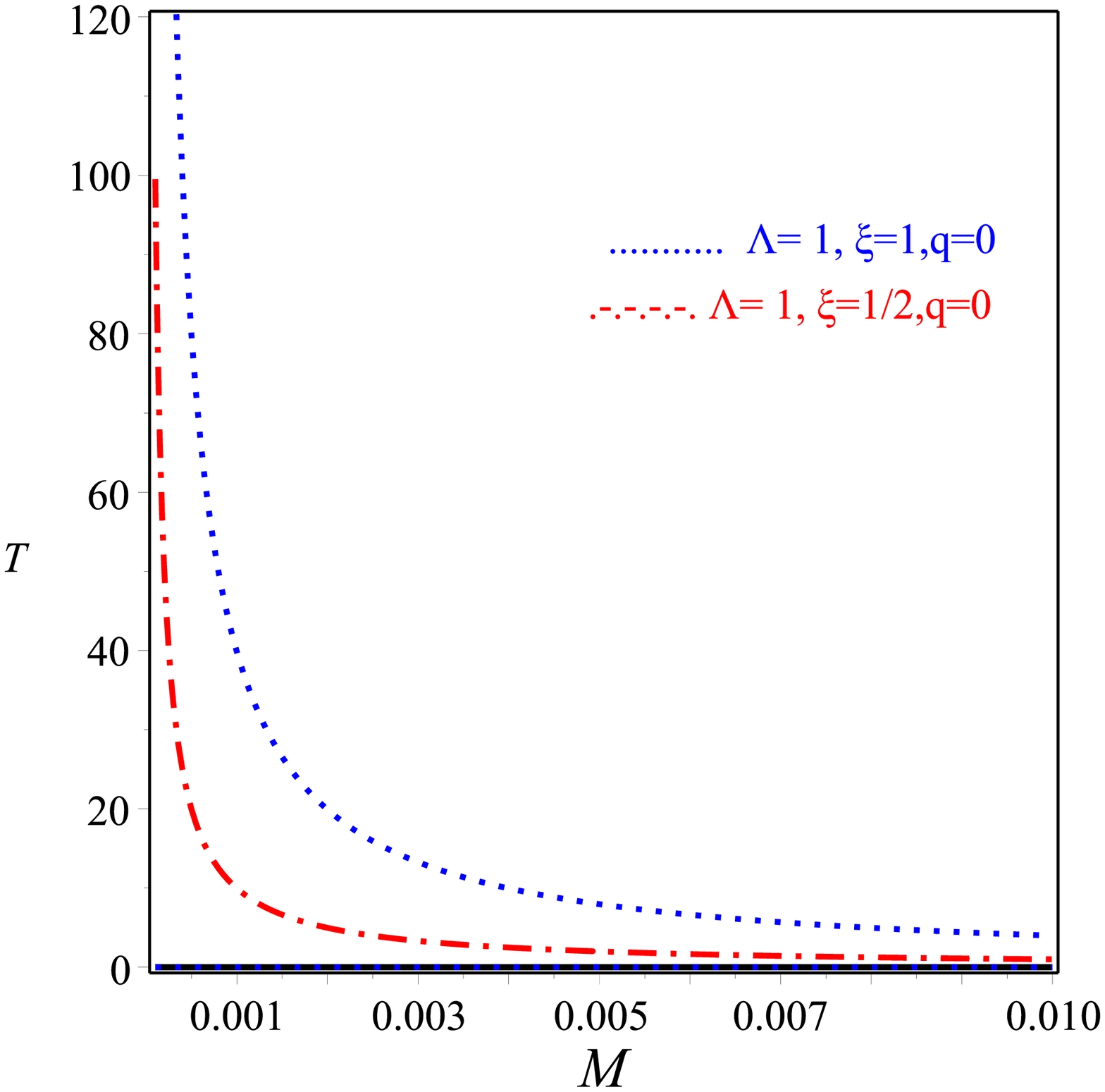}}
\subfigure[~Hawking temperature of BH (\ref{met333})]{\label{fig:Temp1}\includegraphics[scale=0.21]{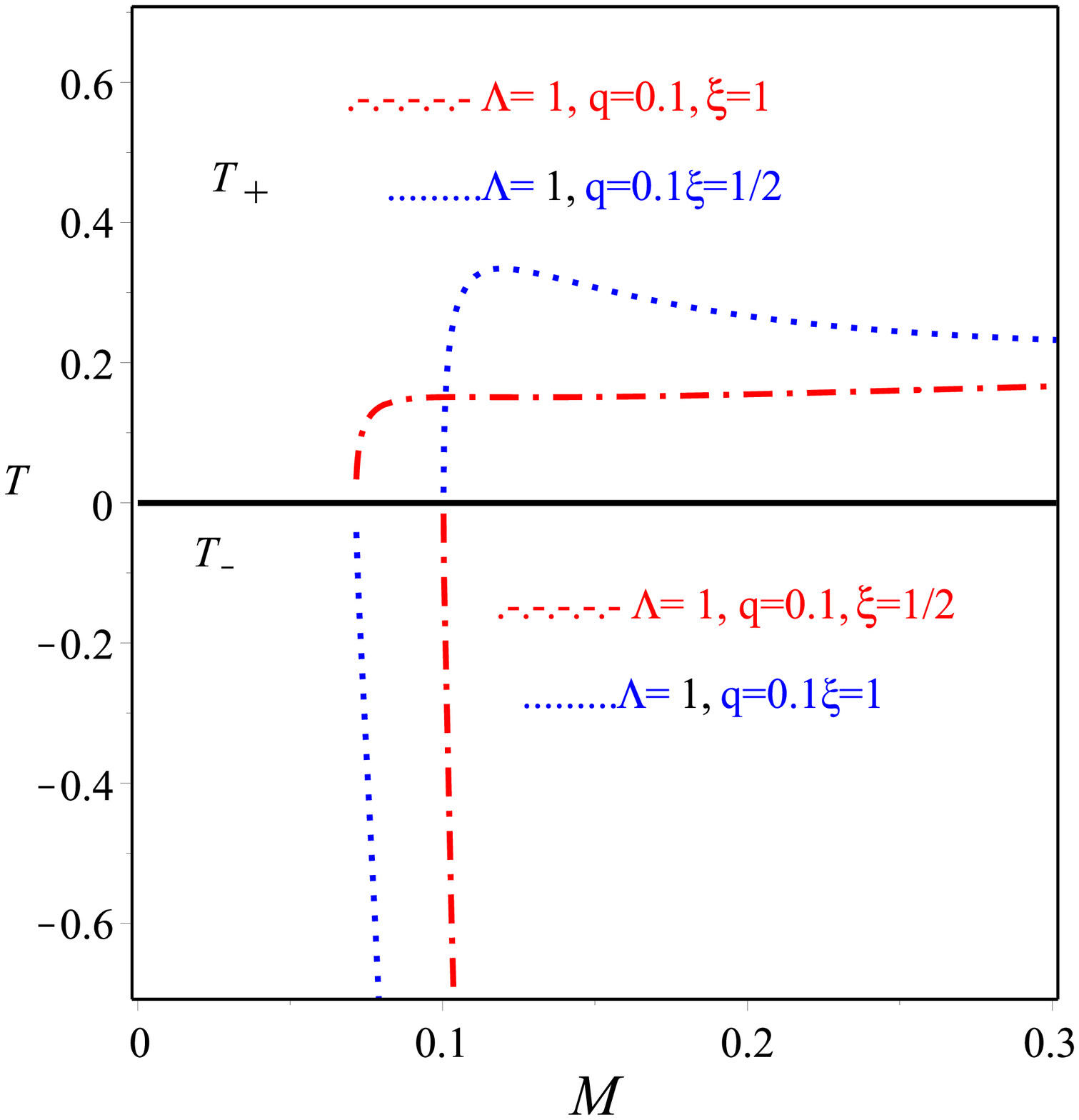}}
\subfigure[~Entropy of BH  (\ref{met33s})]{\label{fig:ent3}\includegraphics[scale=0.21]{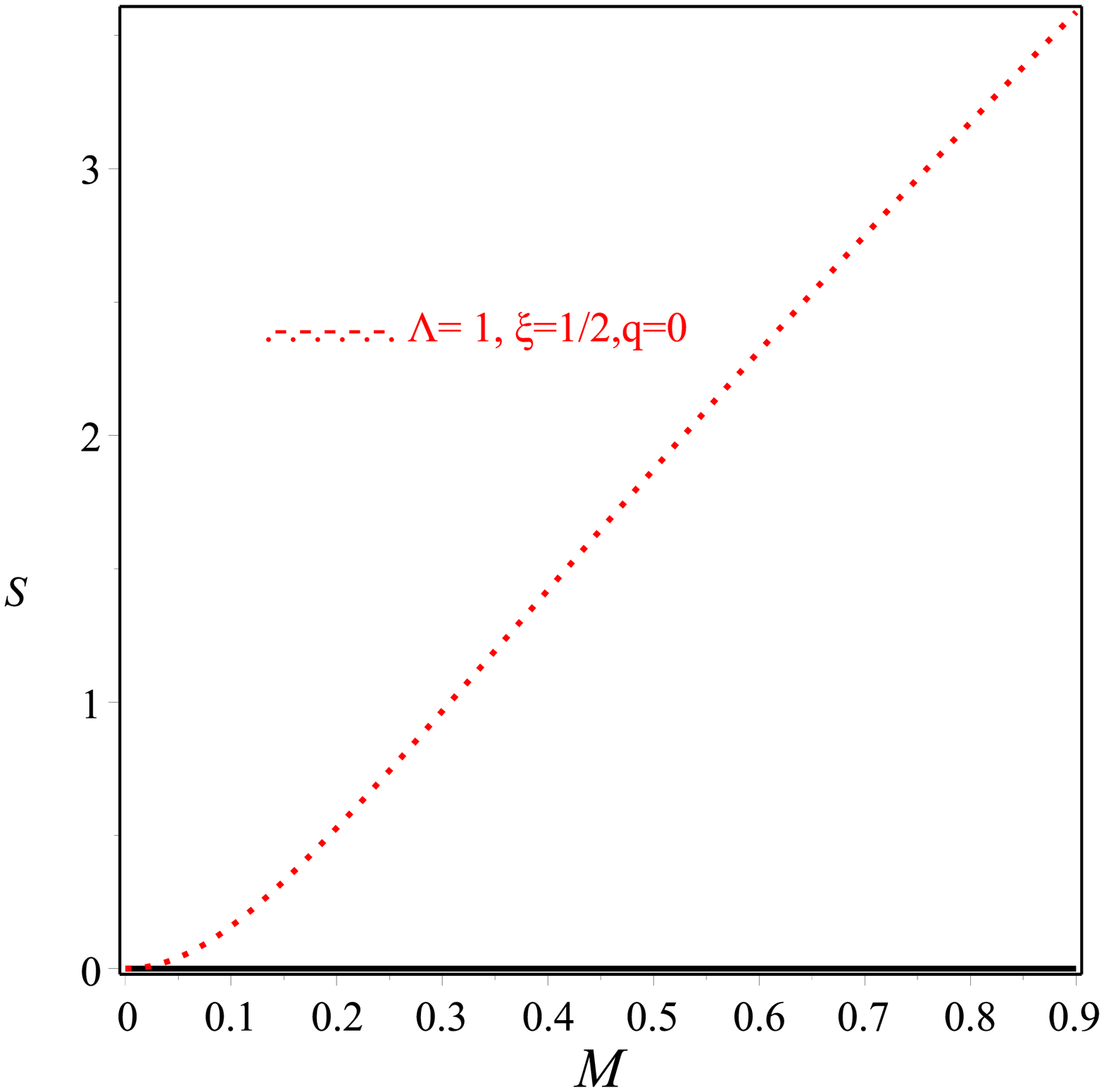}}
\subfigure[~Entropy of BH  (\ref{met333})]{\label{fig:ent1}\includegraphics[scale=0.21]{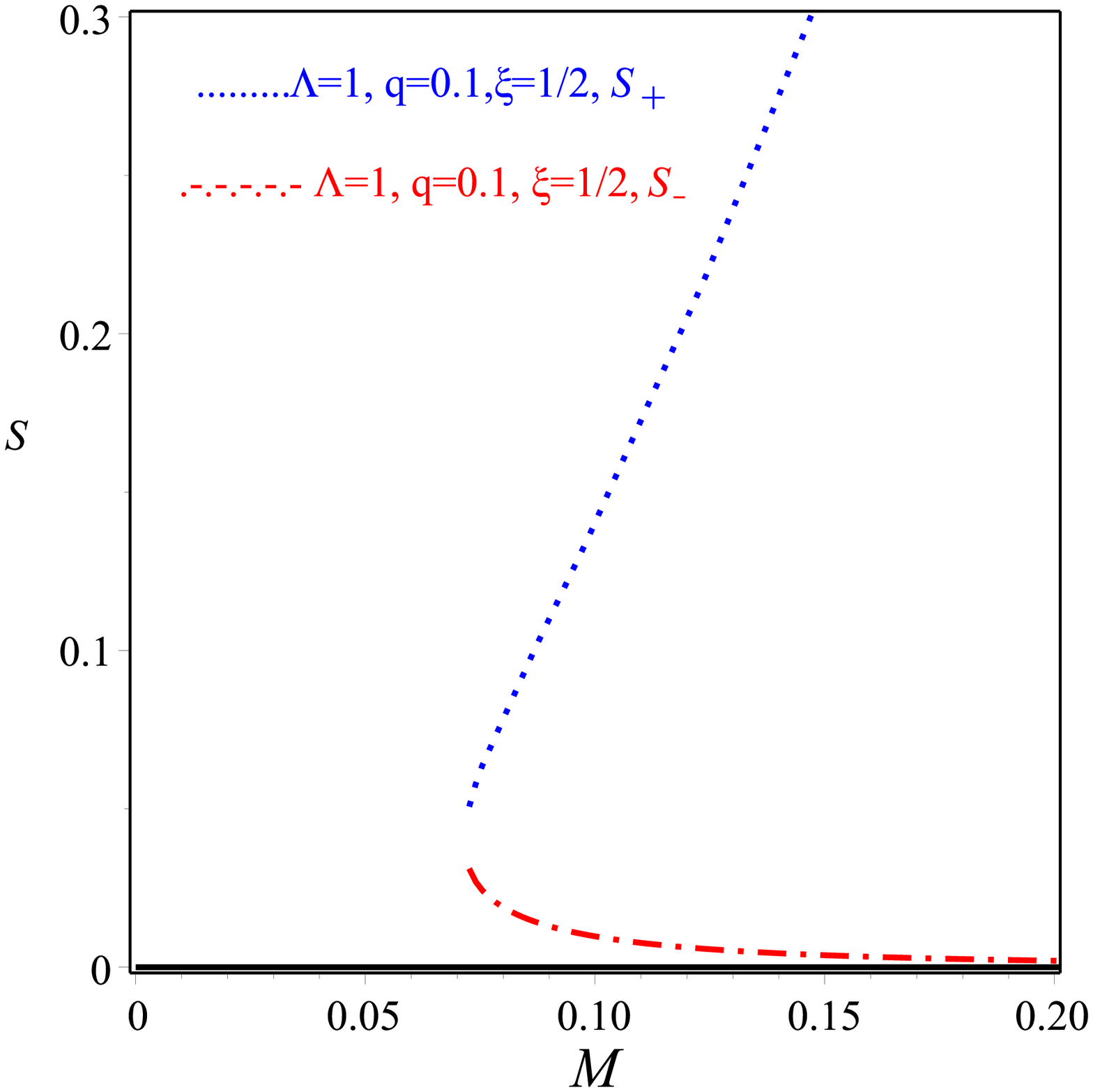}}
\subfigure[~The energy  (\ref{met33s})]{\label{fig:Enr3}\includegraphics[scale=0.21]{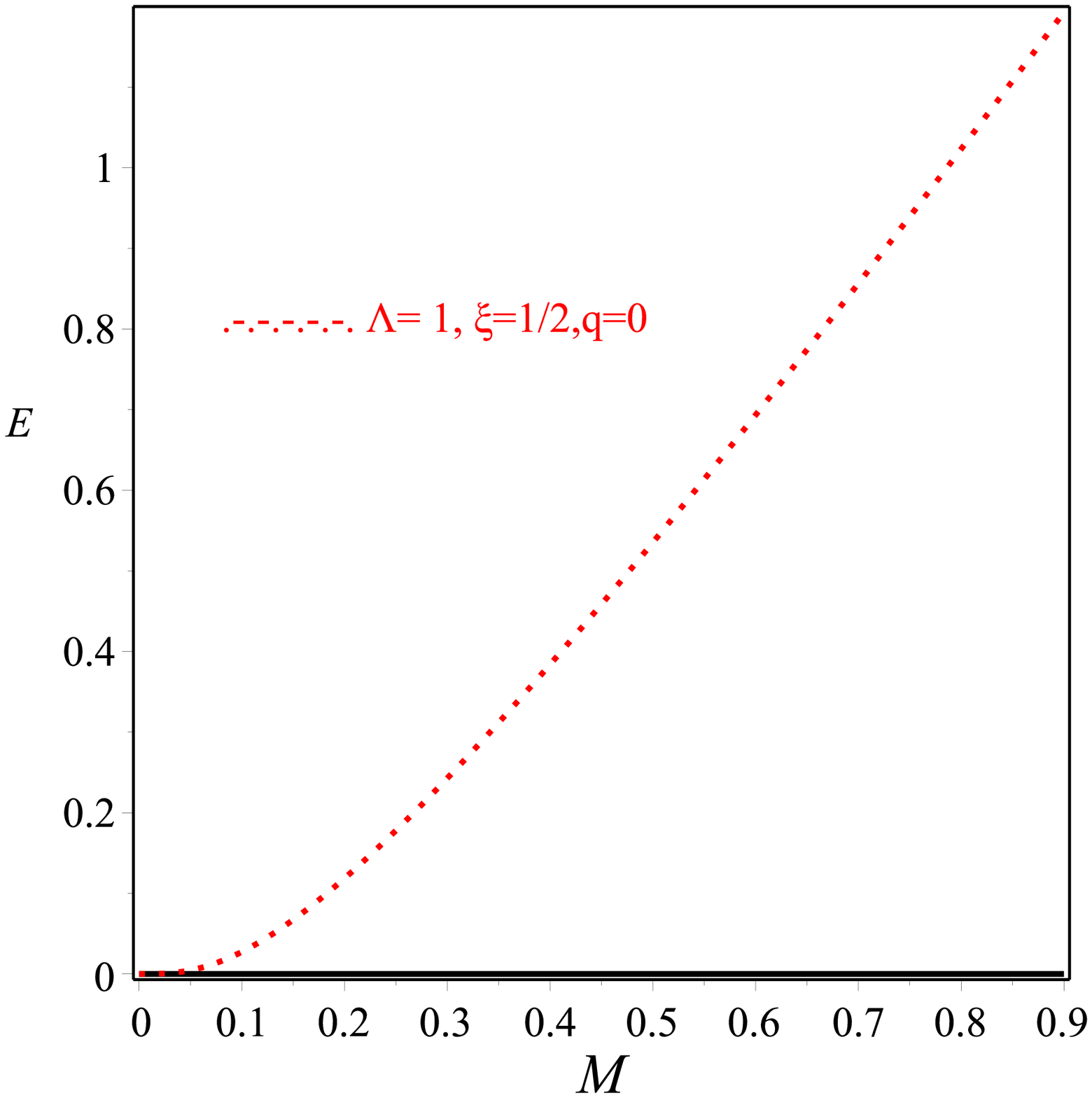}}
\subfigure[~The energy  (\ref{met333})]{\label{fig:Enr}\includegraphics[scale=0.21]{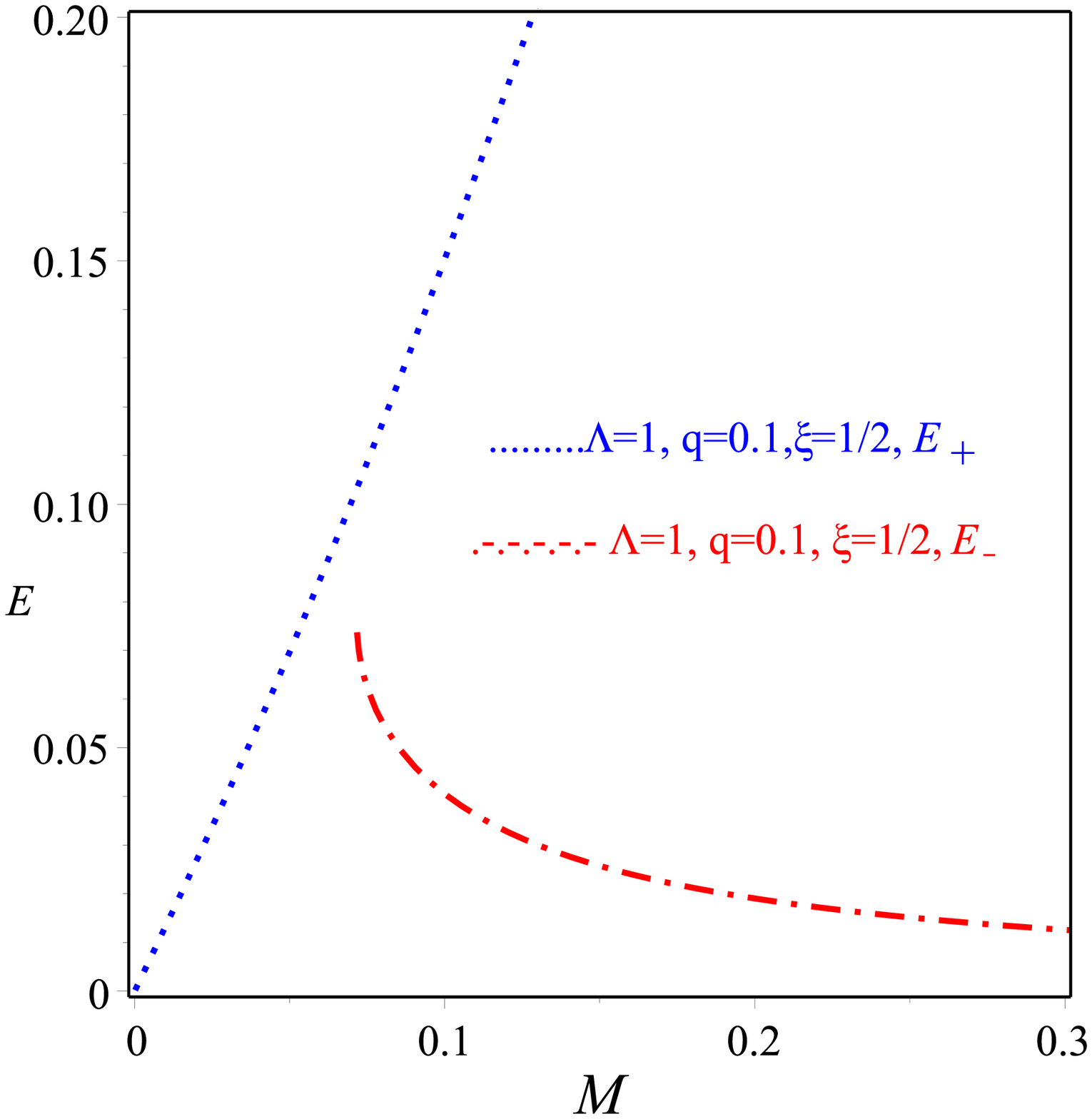}}
\caption[figtopcap]{\small{{Plot of the metric potential,  Hawking temperature and the entropy of Eq.  (\ref{met333}).}}}
\label{Fig:2}
\end{figure}

Solution  (\ref{met333})   is characterized by the  BH mass $M$, the electric field $q$, the cosmological constant $\Lambda$ and  the parameter $\xi_4$ and when the parameter $\xi_4=1$  we get the Reissner-Nordstr\"om  (A)dS space-time. To find the horizons of this BH, (\ref{met333}), we put Eq. $h(r)=0$ which  gives four\footnote{The calculations of horizons, Hawking temperature, entropy and quasi-local energy of the BHs (\ref{met33s}) and (\ref{met333}) are tedious to write them, but we draw their behaviors in Figure  \ref{Fig:2}.}  roots two of them have real values and when $q=0$ we get one real positive root, i.e., \begin{eqnarray}\label{r3} r=\frac{2^{1/3}[(6M\sqrt{\Lambda}+\sqrt{2\xi_4{}^3+36M^2\Lambda})^{2/3}-\xi_4{}^{1/3}]}{2\sqrt{\Lambda}(6M\sqrt{\Lambda}+\sqrt{2\xi_4{}^3+36M^2\Lambda})^{1/3}}\,, \end{eqnarray}
 which is the horizon of the BH (\ref{met33s}).
 The metric potentials of the BHs (\ref{met33s}) and (\ref{met333}) are drawn  in Fig. \ref{Fig:2} \subref{fig:met} and  \subref{fig:met1}  respectively.

Using Eq. (\ref{temp}) we get the Hawking temperature whose form when $q=0$  gives
\begin{eqnarray}\label{T3}
T=\frac{1}{4\pi}\Big[\frac{4M2^{1/3}\Lambda(6M\sqrt{\Lambda}+\sqrt{2\xi_4{}^3+36M^2\Lambda})^{2/3}}{[(6M\sqrt{\Lambda}+\sqrt{2\xi_4{}^3+36M^2\Lambda})^{2/3}
-\xi_4{}^{1/3}]^2}+\frac{2\sqrt{\Lambda}2^{1/3}(6M\sqrt{\Lambda}+\sqrt{2\xi_4{}^3+36M^2\Lambda})^{2/3}
-\xi_4{}^{1/3}]}{3(6M\sqrt{\Lambda}+\sqrt{2\xi_4{}^3+36M^2\Lambda})^{1/3}}\Big]\,.\end{eqnarray}
The behavior of the Hawking temperature of the BHs (\ref{met33s}) and (\ref{met333})  is drawn   in Fig. \ref{Fig:2} \subref{fig:Temp3} and \subref{fig:Temp1} that show  the behavior of the temperature for $\xi_4=1$ and $\xi_4=1/2$ and proves that $T_+>T_-$ for the BH (\ref{met333}). Using Eq. (\ref{ent}) we calculate the entropy   and when $q=0$ we get  the entropy of BH (\ref{met33s})
 whose  behavior  are figured out  in Fig. \ref{Fig:2} \subref{fig:ent3} and  \subref{fig:ent1}.  Using Eq. (\ref{en}) we calculate the quasi-local energy  of the BH (\ref{met333}) and when $q=0$ we get the quasi-local energy  of the BH (\ref{met33s}).  The behavior of these energies are drawn in Fig. \ref{Fig:2} \subref{fig:Enr3} for $q=0$ and in  \ref{Fig:2} \subref{fig:Enr1} for $q\neq 0$.

%
\subsection{Fist law of thermodynamics}\label{S777}
To test if the first law of thermodynamics of the BHs, (\ref{met33}), (\ref{met333f}), (\ref{met33s}) and  (\ref{met333}) are satisfied or not   we are going to use  the formula given in $F(R)$ gravitational theory that has the form \cite{Zheng_2018}
\begin{equation}\label{1st}
dE=TdS-Pdv,
\end{equation}
where $E$, is the quasi-local energy defined by Eq. (\ref{en}), $S$ is Bekenstein-Hawking entropy, $T$ is the Hawking temperature, $P$ is the radial component of the stress-energy tensor that plays the role of thermodynamical pressure, $P=T_r{}^r\mid_{\pm}$, and $V$ is the geometric volume of the space-time. The pressure,  in the context  of $F(R)$ gravitational theory, is defined as \cite{Zheng_2018}
\begin{equation}
P=-\frac{1}{8\pi}\Big\{\frac{{\textit F_R}}{r_{\pm}{}^2}+\frac{1}{2}(f-R{\textit F_R})\Big\}+\frac{1}{4}\Big(\frac{2F}{r_{\pm}}+F'\Big)T\,,\label{11st}
\end{equation}
For the flat space-time given by Eq. (\ref{met33}) we get
\begin{eqnarray} \label{Pr1}
  P_{_{9}}=\frac{12M\xi r(1-\xi)-2r^2+r^2\xi+24M^2(\xi-1)+3Mr}{24rM(\xi-1)}=\frac{5\xi-4}{24\xi(\xi-1)}\,,
\end{eqnarray}
where we have substituted the value of $r$ from Eq. (\ref{r2}) in the case of $q=0$.

Using Eqs. (\ref{T2}) and  (\ref{S2}) in case $q=0$ we get
\begin{eqnarray} \label{Pr3}
T_{_{9}}dS_{_{9}}=\frac{8M(\xi-1)+r}{8r(\xi-1)}=\frac{4\xi^2-4\xi+1}{8(\xi-1)}.
\end{eqnarray}
Finally, using  Eq. (\ref{E2}) in case $q=0$ we get
\begin{eqnarray} \label{Pr4}
dE_{_{9}}=\frac{12M\xi(\xi-1)+2r-\xi r}{24M(\xi-1)}=\frac{6\xi^2(\xi-1)+2-\xi}{12\xi(\xi-1)}.
\end{eqnarray}
Using Eqs. (\ref{Pr1}), (\ref{Pr3}) and (\ref{Pr4}) we can  easily show that the first law given by Eq. (\ref{1st}) is satisfied for the BH (\ref{met33}) when $\xi=1/2$.
Using the same procedure for the BH (\ref{met333f}) we get
\begin{eqnarray} \label{Pr11}
&&P_{\pm(18)}=\frac{4q^2\xi_1{}^2+\xi_1(5M^2-2q^2)+(\pm5\xi_1 M\mp4M)\sqrt{M^2+q^2\xi_1}-4M^2}{24M(M\pm\sqrt{M^2+q^2\xi_1})(\xi_1-1)}\,, \nonumber\\
&&T_{\pm(18)} dS_{\pm(18)}=\frac{[8M\xi_1(1-\xi_1)+M\pm \sqrt{M^2+q^2\xi}][\xi_1q^2+M^2\pm M\sqrt{M^2+q^2\xi_1}]}{8M(M\pm\sqrt{M^2+q^2\xi_1})(\xi_1-1)}\,,\nonumber\\
&&dE_{\pm(18)}=\frac{2M+(\pm 2\mp\xi_1)\sqrt{M^2+q^2\xi_1}+\xi_1M(11-24\xi_1+12\xi_1{}^2)}{24M\xi(\xi_1-1)}\,.
\end{eqnarray}
For  the  BH (\ref{met33s}) we get
\begin{eqnarray} \label{Pr111}
&&P_{(21)}=\frac{12M\xi_2(1-\xi_2)-2r+r\xi_2+48r^2M\Lambda(\xi_2-1)+3M+6\Lambda r^3}{24M(\xi_2-1)}\,, \nonumber\\
&&T_{(21)} dS_{(21)}=\frac{3M+2\Lambda r^3}{24M(\xi_2-1)}\,,\qquad dE_{(21)}=\frac{r(\xi_2-2+4\Lambda r^2)}{24M\xi(\xi_1-1)}\,.
\end{eqnarray}
Substituting the value of $r$ from Eq. (\ref{r3}) and using Eq. (\ref{Pr111}) in (\ref{1st}) we can verify the 1st law of thermodynamics of the BH (\ref{met33s}) when $\xi=1/2$.
 Finally, for the BH (\ref{met333}) we get
\begin{eqnarray} \label{Pr1111}
&&P_{(26)}=\frac{12Mr\xi_4(1-\xi_4)-r^2(2-\xi_4)-48r^3M\Lambda(1-\xi_4)+2r^4\Lambda+3(Mr+q^2)}{24Mr(\xi_1-1)}\,, \nonumber\\
&&T_{(26)} dS_{(26)}=\frac{[8M(1-\xi_4)+r][2\Lambda r^4-3(rM+q^2)]}{24Mr^2(\xi_1-1)}\,,\nonumber\\
&&dE_{(26)}=\frac{12M\xi_4(\xi_4-1)+r(2-\xi_4)+48r^2M\Lambda(1-\xi_4)-4r^3\Lambda}{24M(\xi_1-1)}\,.
\end{eqnarray}

Substituting the values of the positive roots of $h(r)$ given by Eq.  (\ref{met333}) in (\ref{Pr1111})   we can verify the first law of thermodynamics of the BH  (\ref{met333}).
\section{Summary of the main results }\label{S77}
{In this study, we have shown that for a spherically symmetric space-time that has equal metric potentials, i.e., $g_{tt}=\frac{1}{g_{rr}}$ there is only one solution that deviates from GR and  asymptotes as a flat space-time for the specific class of ${\textit  F(R)}$ that has the form ${\textit  F(R)=R-2\alpha\sqrt{R}}$.} Also for the space-time that asymptotes as (A)dS, there is also one BH that deviates from GR BH. To prove the preceding statements we apply the neutral field equation of ${\textit F(R)=R \pm F_1(R)}$  to a spherically symmetric space-time with metric potential has the form $h(r)=\xi-\frac{2M}{r}$ and shows that the only value that the parameter $\xi$ can take is $\xi=1/2$ to get a BH different from GR. We repeat the same technique for the charged case and reach the same conclusion. {\textit It is important to stress on  the fact that some values for $\xi$, except 1/2, give  ${\textit F_1(R)\propto R}$. The reason for this is the fact that  any value of $\xi$ except than one makes Ricci scalar  not vanishing, as is clear from Eq. (\ref{Ricci}), and the form of ${\textit F_1(R)}$ is always $\propto \frac{1}{r^2}$ which makes the form of  ${\textit F(R)}$ always $\propto R.$ The only value  that makes  ${\textit F(R)\neq} R$ is $\xi=1/2$ because in that case ${\textit F_1(R)\propto \frac{1}{r}}$.} The form of  ${\textit F_1(R)}$ that makes the BHs different from GR has the form  ${\textit F_1(R)=\mp\frac{\sqrt{R}}{3M}}$ and the Ricci scalar of these solution takes the form ${\textit R=\frac{1}{r^2}}$. We repeat the same calculations for a spherically symmetric space-time, with metric potential $h(r)=\xi-\frac{2M}{r}+\frac{2\Lambda r^2}{3}$, that asymptotes as (A)dS space-time and use the neutral and charged field equations of  ${\textit F(R)}$ and proof that the only solution that deviates from GR exists only when $\xi=1/2$ and the form of  ${\textit F_1(R)=\mp\frac{\sqrt{\textit R+8\Lambda}}{3M}}$. The Ricci scalar of  this case takes the form  ${\textit R}=\frac{1}{r^2}-8\Lambda$.

To make the picture more clear we calculate some thermodynamic quantities like entropy, Hawking temperature, quasi-local energy and Gibbs free energy for the two space-times that asymptotes as flat  and (A)dS  space-times. We show that for both space-times that the thermodynamic quantities for the parameter $\xi=1/2$   are physically acceptable  consistent with the literature.  Thus the case where $\xi=1/2$ that deviates from GR is physically acceptable from the viewpoint of thermodynamics. Finally, we showed by detail  calculations that all the derived BHs,  (\ref{met33}), (\ref{met333f}), (\ref{met33s}) and  (\ref{met333}), satisfy   the first law   of  thermodynamics given by  Eq. (\ref{1st}).

{ It is well-known that for any null geodesic in the exterior region of the Schwarzschild metric the null geodesic equations are given by \cite{Claudel:2000yi}
\begin{equation}\label{EHT}
\frac{d^2r}{d\eta^2}=(r-3M)\Biggl[\left(\frac{d\theta}{d\eta}\right)^2+\sin^2\theta\left(\frac{d\phi}{d\eta}\right)^2\Biggr]\,,
\end{equation}
where $\eta$ is the affine parameter along the geodesic. The R.H.S. of Eq. (\ref{EHT}) has a positive value for $r>3M$ and has a negative value when $3M>r>2M$.  For the BH (\ref{met2}), the null geodesic equations are given by
\begin{equation}\label{EHT1}
\frac{d^2r}{d\eta^2}=\frac{1}{2}(r-6M)\Biggl[\left(\frac{d\theta}{d\eta}\right)^2+\sin^2\theta\left(\frac{d\phi}{d\eta}\right)^2\Biggr]\,.
\end{equation}
The R.H.S. of Eq. (\ref{EHT1}) has a positive value when $r>6M$ and negative value when $6M>r>4M$. Therefore, for any future endless null geodesic in the maximally extended  space-time given by Eq. (\ref{met2}) starting at some point with $6M<r$ and initially directed outwards, in the sense
that $\frac{dr}{d\eta}$ is initially positive, will continue outwards and escape to infinity. Any future endless null geodesic in the maximally extended space-time given by Eq. (\ref{met2})  starting at some point with $ r: 6M > r >4M$  and initially directed inwards, in the sense that $\frac{dr}{d\eta}$ is initially negative will continue inwards and fall into the black
hole. The hypersurface ${r = 6M}$, known as the  photon sphere of the space-time (\ref{met2}), thus distinguishes the borderline between these two types of behavior; any null geodesic
starting at some point of the photon sphere and initially tangent to the photon sphere will remain in the photon sphere;  for more detail,  see \cite{doi:10.1098/rspa.1961.0142,doi:10.1098/rspa.1959.0015}}.

To conclude, we have shown that in the frame of ${\textit F(R)=R+F_1(R)}$ there is only a unique spherically symmetric BH, form the line-element which has equal metric potential,  that is different from GR which asymptotes as flat or (A)dS space-times. Is this result valid for the spherically symmetric line-element that has unequal metric potentials? If yes is its thermodynamic  quantities are acceptable? These questions will be answered elsewhere.

\section*{Acknowledgments}
The author acknowledges the anonymous Referee for improving the presentation of the manuscript.

%

\end{document}